\documentclass[review]{elsarticle}

\usepackage[utf8]{inputenc}
\usepackage{float}
\usepackage{amsmath}
\usepackage{amssymb}

\numberwithin{equation}{section}
\newtheorem{exmp}{Example}[section]

\makeatother

\graphicspath{ {./img_color/} }
\newtheorem{defn}{Definition}[section]

\begin{document}

\title{Predicting multidimensional distributive properties of hyperbranched
polymer resulting from $AB_{2}$ polymerization with substitution,
cyclization and shielding }

 \author[uva]{I.~Kryven\corref{cor1}}
 \ead{i.kryven@uva.nl}
 \author[uva]{P.~D.~Iedema}
 \ead{p.d.iedema@uva.nl}
\cortext[cor1]{Corresponding author}
\address[uva]{University of Amsterdam, Science Park 904, \\1098 XH Amsterdam, The Netherlands}

\begin{abstract}
A deterministic mathematical model for the polymerization of hyperbranched
molecules accounting for substitution, cyclization, and shielding effect
has been developed as a system of nonlinear population balances. The
solution obtained by a novel approximation method shows perfect agreement
with the analytical solution in limiting cases and provides, for the
first time in this class of polymerization problems, full multidimensional
results.
\end{abstract}


\maketitle
\tableofcontents{}
\section*{Introduction}

In this paper the mathematical modelling of hyperbranched polymers
based on monomers of the $AB_{2}$ type is taken at hand, allowing
the calculation of various microstructural properties, using a numerical-mathematical
scheme based on a graph-theoretical description of branching topology.
. In previous work, scalar quantities like average molecular weight,
polydispersity, degree of branching, or gel point have been successfully
predicted. These results were mainly achieved either by statistical
theories of Stockmayer \cite{Stockmayer1943,Stockmayer1944} and Flory
\cite{Flory1952} or, later, by stochastic simulations \cite{jo2001}
and a generating functions approach \cite{Galina2002,Cheng2002,Zhou2006}.
Most theoretical models that describe the polymerization of $AB_{2}$
monomer are based on the assumption that all functional groups of
the same type are equally reactive, and react independently of one
another. In other words, all $B$ groups in the $AB_{2}$ monomer
units in the polymer have the same reactivity. Yet, the equal reactivity
is questioned if a \emph{substitution} effect is taken into account
\cite{Cheng2002}.

As regards the $AB_{2}$ problem with substitution, significant progress
has been achieved in studies by Galina\cite{Galina2002,Galina2002b},
Cheng\cite{Cheng2002} and Zhou\cite{Zhou2006,Zhou2011}. Here, the behaviour
in time of the degree of branching\cite{Galina2002,Cheng2002}, the
dendritic and linear units\cite{Cheng2002} and the fraction of cyclized
molecules\cite{Galina2002} has been obtained. In the work by Galina\cite{Galina2002}
and Cheng \cite{Cheng2002} the average microstructural properties
have been computed using the generating functions approach. More recently,
the 2-dimensional chain-length/degree-of-branching distribution was
obtained in the form of an analytical expression by Zhou\cite{Zhou2006}
under the assumption that no cyclization reaction takes place. The
intramolecular reaction of cyclization has indeed been considered
in \cite{Galina2002}. In view of its interesting gelation behaviour, the $AB_{3}$ system has been investigated using Monte
Carlo simulation by Somvarsky et al. \cite{Somvarsky1998} Rather than substitution they
were interested in the effect of \emph{shielding} of monomer groups,
leading to a shift of the gel point to higher conversion. They deal
with the steric excluded volume effect by introducing so-called exponent
\emph{kernels}, thereby mimicking non equal accessibility of reactive
functional groups. In the present paper we show that our newly developed
numerical method can be easily adapted to capture the shielding effect in $AB_2$ system employing an exponent kernel. 

The new computational techniques that we will present in this paper
are another contribution to the area of growing interest that is formed
by the search after multidimensional distributive properties, where
we previously treated polymer modification \cite{Kryven2012,kryven2013}
and crosslinking polymerization \cite{Iedema2012,Kryven2012,kryven2013}.
In the realm of this trend here we will show how multidimensional
distributive properties can be recovered for the $AB_{2}$ polymerization
system with substitution, cyclization, and shielding. The following
reaction mechanism is considered: 
\[
\begin{array}{l}
P_{x,y}+P_{x',y'}\stackrel{x K_t}{\longrightarrow}P_{x+x'-1,y+y'+1};\\
P_{x,y}+P_{x',y'}\stackrel{y \rho K_t}{\longrightarrow}P_{x+x',y+y'-1};\\
P_{x,y}\stackrel{x \lambda K_t}{\longrightarrow}C_{x-1,y};\\
P_{x,y}\stackrel{y \rho \lambda K_t}{\longrightarrow}C_{x,y-1};\\
P_{x,y}+C_{x',y'}\stackrel{x  K_t}{\longrightarrow}C_{x+x'-1,y+y'+1};\\
P_{x,y}+C_{x',y'}\stackrel{y\rho  K_t}{\longrightarrow}C_{x+x',y+y'-1},
\end{array}
\]
where $x$ is a number of terminal units, and $y$ is a number of liner units in the hyperbranched polymer. The first two lines correspond to the polymerization reaction
for substituted and non substituted groups. The third and forth lines
denote the intramolecular reaction of cyclization. The fifth and sixth
lines denote the reaction between cyclized and non-cyclized molecules. The rate coefficients are not constant, and depend on $x$, $y$, chemical substitution ratio $\rho$, cyclisation factor $\lambda$ and scaling constant $K_t$. 
Note, that we do not yet explicitly refer to shielding in this formulation.

The paper is organised as follows. \\
 \emph{Firstly}, we discuss the main concepts of \emph{graph theory}
that has been used to describe molecular topologies of hyperbranched
$AB_{2}$ polymer. Graph theory is not only an important tool to build
the mathematical model that is going to be solved. It also helps us
understanding the connections between different distributive properties
and, in addition, it allows expressing one property in terms of the
other at the post-processing stage. \emph{Secondly}, we show how the
rules of topology evolution as induced by the chemical reactions,
may be transformed into mathematical balance equations. The non-linear
integro-differential equations will be discretized and solved numerically
employing linearisation and projection methods. \\
 \emph{Thirdly}, we demonstrate how the 'raw' numerical result, a
three-dimensional distribution, may be post-processed in order to
retrieve different distributive properties. The average properties
and the chain length distribution obtained from the model, are shown
to perfectly agree with findings of earlier investigations \cite{Galina2002,Cheng2002,Zhou2006}
for the case without cyclization. Furthermore, we show the first
ever full multidimensional distributive properties for systems
that experience \emph{both} substitution and cyclization simultaneously.
\\
 \emph{Fourthly}, we show how the designed numerical method can be
adapted to capture the steric excluded volume effect by using the
exponent kernels - the first time this concept is realized in a deterministic
manner.

\section{Graph representation for hyperbranched polymers}

\label{sec:Graphs} 
\begin{figure}[h]
\center \includegraphics[width=0.6\textwidth]{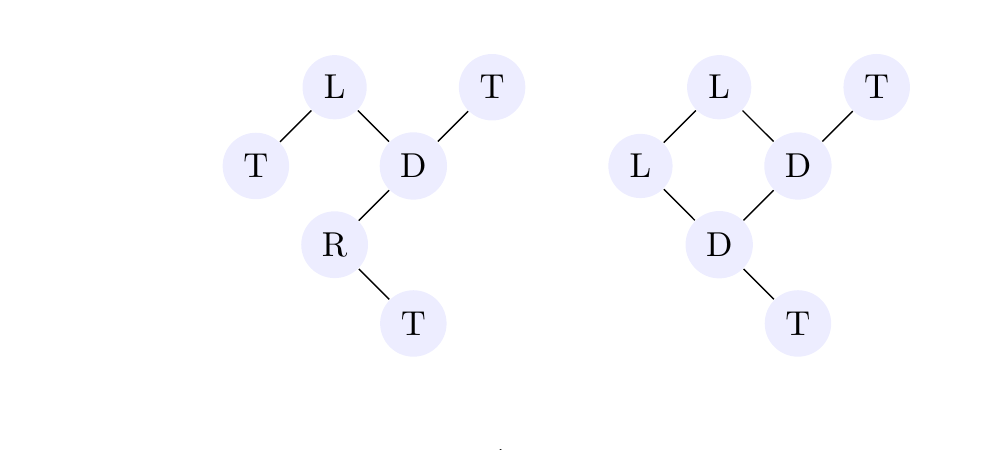} \caption{The graph on the left has a tree topology with root $R$, while the
graph on the right has a cycle. The length of the cycle is equal to
4. }

\label{fig:tree} 
\end{figure}

The topology of a hyperbranched polymer is now
described in terms of \emph{graph theory}. According to this approach,
each monomer in the polymer of the $AB_{2}$ type has one \emph{node}
associated with it, while the chemical bonds between the monomers
are represented by \emph{edges}. A single polymer molecule is represented
by a connected graph $G$ with nodes of maximum \emph{degree} three
and at most one cycle. Moreover, the structure of a non-cyclized polymer
molecule does not contain cycles at all, and is therefore called a
\emph{rooted binary tree}. In the present case of $AB_{2}$ a cyclized
polymer graph has exactly one cycle, see Figure~\ref{fig:tree}.
Now, it is useful to distinguish monomers according to their position
in the topology graph. Table~\ref{tab:dict} shows the correspondence
between monomer position, graph theory terms, and commonly used short
names \cite{Galina2002,Cheng2002}.

\begin{table}[h]
\begin{tabular}{p{5cm}|c|c}
Monomer  & Graph theory term  & Short name \tabularnewline
\hline 
two reacted $B$ groups, and unreacted $A$  & root node  & \emph{root}, $R$ \tabularnewline
two unreacted $B$ groups, $A$ reacted  & node of degree one  & \emph{terminal unit}, $T$ \tabularnewline
one reacted and one unreacted group $B$, $A$ reacted  & node of degree two  & \emph{linear unit}, $L$ \tabularnewline
all $A$ and $B$ groups reacted  & node of degree three  & \emph{dendritic unit}, $D$ \tabularnewline
polymer with no unreacted $A$ groups  & graph with a cycle  & cycle, $C$ \tabularnewline
\end{tabular}\caption{A graph theoretical frame of reference of $AB_{2}$ polymer molecules.}

\label{tab:dict} 
\end{table}

\begin{figure}[h]
\center \includegraphics[width=0.8\textwidth]{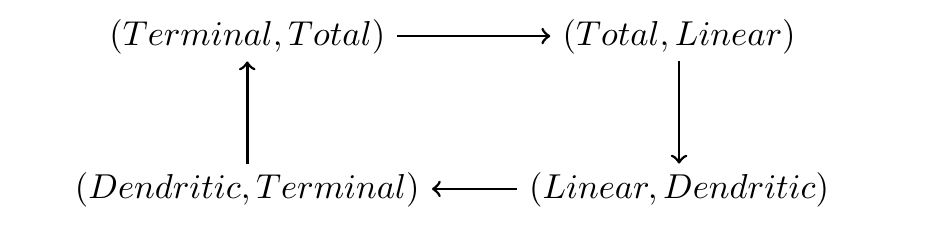} \caption{By knowing two of four parameters it is possible to recover the other
two.}

\label{fig:square} 
\end{figure}

The graph representation brings a lot of advantages along, for instance,
using basic equalities from graph theory \cite{diestel2000} one immediately
infers the connection between the parameters given in Table~\ref{tab:dict}
take place. In fact, it suffices knowing only two of the parameters
in order to reconstruct the others, see Figure~\ref{fig:square}.

Indeed, the dependence of the total number of units on the number
of terminal and linear units is given by the rather trivial expression,
\begin{equation}
N=2T+L+C-1.\label{eq:trivial}
\end{equation}

Now, the $AB_{2}$ polymerization process leading to branched topologies,
can be viewed as a specific \emph{random graph process} \cite{Erdos}
that starts with $n$ nodes and no edges, and subsequently at each
step connects certain nodes according to the reactions taking place.

\section{Reaction mechanisms and mathematical model}

Let $\bar{G}=\{G\}$ be a collection of graphs (sometimes referred
to as a \emph{forest} \cite{diestel2000}) with, possibly repeating,
components of the form $G$. Hence, $\bar{G}$ is a \emph{multiset}.
Employing this notation, we refer to $G$ as to a single polymer molecule,
while $s\in\bar{G}$ represents all molecules instantaneously present
in the system, and $\bar{G}$ all possible molecules that may be generated
with the $AB_{2}$ polymerization process.

Following the conventional approach for $AB_{2}$ systems with substitution,
two polymerization reactions\cite{Galina2002,Cheng2002} and a cyclization\cite{Galina2002}
reaction are considered. Recalling $x,y$ denotes amount of terminal and linear units, the interpretation of the reactions in terms of the random
graph process emerges as: 
\begin{itemize}
\item two arbitrary tree-components of $\bar{G}$ join together by connecting
the root $R$ from one component and a terminal unit $T$ from another
at rate $K_t$, 
\[
P_{x,y}+P_{x',y'}\rightarrow P_{x+x'-1,y+y'+1};
\]

\item two arbitrary tree-components of $\bar{G}$ (at least one of two is
not strictly binary) join together by connecting the root $R$ from
one tree and a linear unit $L$ from another at rate $\rho K_t$,
\[
P_{x,y}+P_{x',y'}\rightarrow P_{x+x',y+y'-1};
\]

\item an arbitrary tree component of $\bar{G}$ receives a cycle by connecting
the root $R$ with a terminal unit $T$ at rate $\lambda K_t$ or
by connecting root $R$ and linear unit $L$ at rate $\rho K_t$,

\[
P_{x,y}\rightarrow C_{x-1,y},
\]
\[
P_{x,y}\rightarrow C_{x,y-1};
\]

\item two arbitrary tree-components of $\bar{G}$, of which one possesses
a cycle, join together by connecting a root $R$ from one molecule
and a terminal unit $T$ from another at rate $K_t$; or join by
connecting root the $R$ from one tree and a linear unit $L$ from
another at rate $\rho K_t$,

\[
P_{x,y}+C_{x',y'}\rightarrow P_{x+x'-1,y+y'+1},
\]
\[
P_{x,y}+C_{x',y'}\rightarrow P_{x+x',y+y'-1}.
\]

\end{itemize}
Suppose we have a system of polymers $s$. Let each graph $g\in G$
has a number $0\leq F_{s}(g)\leq1$ associated with it that corresponds
to the relative frequency of occurrence of graph $g$ within the system
$s\in\bar{G}$. In other words, for each selected topology $g$, $F_{s}(g)$
tells how probable it is finding $g$ within a particular system $s$.
As $g$ contains full information on molecular topology, it is important
to note that the whole system $s$, consisting of a large number of
molecules, would have to be represented by an ensemble of the above-described
graphs, which would require a huge amount of data to store. Thus,
we will rather distinguish topologies only by three parameters ($T,L,C$),
mapping $G\rightarrow\Omega\times\{0,1\},\;\Omega=[0,\infty)^{2}$.
Here $\Omega$ denotes the domain of all possible values for parameters
$(T,L)$. On the other hand, we expect $s$ to be exclusively dependent
on time by considering a function $s(t)$. This reduces the original
measure $F_{s(t)}$ to 
\begin{equation}
f_{c}(x,y,t)\in W, \; c\in\{0,1\},
\end{equation}
which denotes the relative frequency of occurrence of a connected
component with a number of terminal units between $x$ and $x+dx$,
a number of linear units between $y$ and $y+dy$, and $c$ cycles
in the polymerization system at time $t$. We refer to $W=L^{2}(\Omega)\times C[1,\infty),$ as a short way to describe mathematical properties of the distribution $f(x,y,t)$.
Namely, $L^{2}(\Omega)$ suggests $f_{c}(x,y,t)$ is integrable over the
first two variables $x,y$; $C[1,\infty]$ mark it is continuous in time, $t$. We will deliberately use this shortening in the paper. For instance, when referring to a mapping $W \rightarrow W$, a law that sets correspondence between two distributions from $W$; a mapping $W^2\rightarrow W$ - correspondence between a pair of distributions and another distribution from $W$; and $W\rightarrow \mathbb{R}$ - correspondence between a distribution for $W$ and a real number.

Passing to the limit $dx,dy\rightarrow0$ the evolution of $f_{c}(x,y,t)$
with time may be described with a system of differential equations.
Two equations are derived, for cyclized polymers $c=1$, and for non-cyclized
ones, $c=0$, 
\begin{equation}
\frac{\partial f_{c}(x,y,t)}{\partial t}=(\mathcal{L}_{c}f)(x,y,t),\; c=0,1\label{eq:theeq}
\end{equation}
subject to the initial conditions 
\begin{equation}
\left\{ \begin{array}{l}
f_{0}(x,y,0)=\delta(x-1)\delta(y),\\
f_{1}(x,y,0)=0,
\end{array}\right.\label{eq:initial_conditions}
\end{equation}

where $\delta$ is a delta function and operator $\mathcal{L}:W\rightarrow W$
implements the effect of the reactions listed at the beginning of
this paragraph on the distribution $f_{c}$. The operator fully determines
the dynamics of the system and, as it will be shown further on, has
a non-linear structure. Now, we start with a set of basic operators,
typical for polymerization, that also act on $f_{c}$ and eventually
will serve as building blocks to define a more complicated structure
of $\mathcal{L}$. 
\begin{defn}
The \emph{shift operator} $T_{x_{s},y_{s}}\in W\rightarrow W$ takes a function
$f\in W$ to its translation $T_{x_{s},y_{s}}f$, 
\begin{equation}
T_{x_{s},y_{S}}f(x,y,t)=f(x-x_{s},y-y_{s},t),\;(x_{s},y_{s})\in\Omega.\label{eq:shift}
\end{equation}
\end{defn}
We normally associate the operator \eqref{eq:shift} with propagation-like
reaction mechanisms, where at each firing of the reaction one node
is added to the connected component $G$. \begin{exmp}\label{ex:shift}
Let us consider a simple propagation reaction $P_{1,y}+M\stackrel{K_{t}}{\longrightarrow}P_{1,y+1}$,
meaning that the lengths of the linear chains $P_{1,y}$ are growing
at constant rate $K_t$. The corresponding differential equation has
the form 
\[
\frac{\partial{}}{\partial t}f(x,y,t)=K_{t}\Big(T_{0,1}f(x,y,t)-f(x,y,t)\Big).
\]
\end{exmp} The differential equation from Example~\ref{ex:shift}
may be viewed as a partial case of a chemical master equation \cite{gillespie2000}.
Note besides that any chemical master equation may be represented
as a linear combination of shift operators \eqref{eq:shift}.

\begin{defn}
\emph{Integral operators} $\mu,\mu_{x},\mu_{y}:W\rightarrow\mathbb{R}$,
take a function $f$ to its partial moments $\mu f,\mu_{x}f,\mu_{y}f$:
\begin{equation}
\begin{array}{rcr}
(\mu f)(t) & = & \int\limits _{\Omega}f(x,y,t)\, dxdy,\\
(\mu_{x}f)(t) & = & \int\limits _{\Omega}xf(x,y,t)\, dxdy,\\
(\mu_{y}f)(t) & = & \int\limits _{\Omega}yf(x,y,t)\, dxdy,
\end{array}\label{eq:moments}
\end{equation}
\end{defn}
\begin{exmp} Let $f_{c}(x,y,t)\in W$ be a known evolution of the
topology for an $AB_{2}$ polymerization system as described in Section~\ref{sec:Graphs}.
Then, as a direct consequence of the equality \eqref{eq:trivial},
the total mass time profile for the non-cyclized polymers ($c=0$)
is given by 
\[
m(t)=(2\mu_{x}+\mu_{y}-\mu)f_{0}(x,y,t).
\]
\end{exmp}

\begin{defn}
\emph{Convolution} is an operation that takes two arguments $f,g\in W$
to a product $f*g$ defined as 
\begin{equation}
f(x,y,t)*g(x,y,t)=\int\limits _{0}^{x}\int\limits _{0}^{y}f(\xi,\eta,t)g(x-\xi,y-\eta,t)\, d\xi d\eta\label{eq:convoluton}
\end{equation}
\end{defn}
It may easily be checked that the convolution is a \emph{symmetrical}
$f*g=g*f$, and a bilinear $a\cdot f*g=a\cdot(f*g),\; a\in\mathbb{R}$
operation. Note, that the operation exhibits special behaviour regarding
the 0-moment: 
\begin{equation}
\mu(f*g)=\mu f\cdot\mu g
\end{equation}
The convolution is an important tool when modelling step-growth polymerization
\cite{Kryven2012}. Firing of such a reaction in terms of a random
graph process means that any two arbitrary disconnected components
from $\bar{G}$ become connected by a new edge. This is illustrated
by the following example. \begin{exmp}\label{ex:convolution} Let
us consider a polymerization reaction described by the reaction equation
$P_{x,y}+P_{x',y'}\stackrel{K_{t}}{\longrightarrow}P_{x+x',y+y'}$, where
the reaction rate is not dependent on $x,y,x',y'$. In contrast to to
Example~\eqref{ex:shift} there are multiple choices to select the
reactants, so $P_{x,y}$ is produced for a fixed combination $x,y$.
The corresponding differential equation has is formulated as, 
\[
\frac{\partial{}}{\partial t}f(x,y,t)=K_{t}\Big(f(x,y,t)*f(x,y,t)-f(x,y,t)\cdot\mu f(x,y,t)\Big).
\]
\end{exmp} The differential equation presented in Example~\eqref{ex:convolution}
can be also viewed as a \emph{Smoluchowski coagulation equation} \cite{smoluchowski}.

Now, we are ready to define $\mathcal{L}(f_{c}(x,y,t)),\; c=0,1,$ as
appearing in \eqref{eq:theeq}. In analogy to the ideas expressed
in Examples~\ref{ex:shift}-\ref{ex:convolution} we assemble the
operator $\mathcal{L}_{0,1}$, 
\begin{align}
\mathcal{L}_{0}(f_{c})=\; & K_t\bigg(T_{0,1}\, f_{0}*xf_{0}-\big(\mu_{x}f_{0}+\mu f_{0}\, x\big)\cdot f_{0}-\lambda xf_{0}\bigg)+{}\notag\label{eq:operator_L0}\\
{} & \rho K_t\bigg(T_{1,-1}\, f_{0}*yf_{0}-\big(\mu_{y}f_{0}+\mu f_{0}\, y\big)\cdot f_{0}-\lambda yf_{0}\bigg);\\
\mathcal{L}_{1}(f_{c})=\; & K_t(T_{0,1}xf_{1}*f_{0}-xf_{1}\mu f_{0})+\rho K_t(T_{1,-1}yf_{1}*f_{0}-yf_{1}\mu f_{0})+\lambda K_t\big(xf_{0}+\rho yf_{0}\big).\label{eq:operator_L1}
\end{align}
Here as before, the subscript denotes the balances for cyclized $1$
and non-cyclized $0$ molecules. The equation \eqref{eq:operator_L0}
consists of two similar lines that correspond to $R+T$ and $R+L$
reactions. Here, $\rho>0$ is the chemical substitution factor. As
similar to Example~\ref{ex:convolution} $T_{0,1}f_{0}*xf_{0}$ denotes
the production term. Here, however, the weight $x$ has been used
as the reaction rate is proportional to the amount of terminal units.
The shift operator $T_{0,1}$ was employed, since with every reaction
firing, one extra linear unit appears. The consumption term $-\big(\mu_{x}f_{0}+\mu f_{0}\, x\big)\cdot f_{0}$
shows that each polymer might participate in the reaction by contributing
either an $R$ unit at rate $\mu f_{0}\cdot xf_{0}$ or a $T$ unit
at rate $\mu_{x}f_{0}\cdot f_{0}$. Finally, the cyclization $R+T$
occurs at rate $\lambda K_txf_{0}$, where $\lambda\geq0$ is a
constant factor. The second line of \eqref{eq:operator_L0} may be
analyzed in an analogous manner with as a major difference that it
denotes the substituted reaction $R+L$ this time.

It may be seen that the balance equation for components without cycles
\eqref{eq:operator_L0} is presented in a closed form, so the differential
equation for $f_{0}$ may be solved independently of $f_{1}$. In
the case $f_{0}$ is known, $f_{1}$ may be resolved by integration
\begin{equation}
\frac{\partial f_{1}(x,y,t)}{\partial t}=\mathcal{L}_{1}(f_{c}(x,y,t)).\label{eq:f1}
\end{equation}
For this reason we subsequently consider two differential equations
instead of the full system \eqref{eq:theeq}. The numerical approach
is similar for the cases and will first be introduced for $\mathcal{L}_{0}$.

\subsection{Numerical treatment}

We start with the time discretization. In view of the non-linear nature
of \eqref{eq:theeq}, it is important to employ an implicit time integration
scheme, also known as Rothe's method. Let $\hat{f}(x,y,t_{k})$ be
an approximation to $f_{0}$, the solution of \eqref{eq:theeq}. Then,
an implicit first order time integration scheme is given by 
\[
\hat{f}(x,y,t_{k+1})=\hat{f}(x,y,t_{k})+\tau_{k}\mathcal{L}_{0}\hat{f}(x,y,t_{k+1}),\; k=0,1,2,\dots
\]
or in a more compact way, 
\begin{equation}
(\tau_{k}\mathcal{L}_{0}-I)\cdot\hat{f}(x,y,t_{k+1})+\hat{f}(x,y,t_{k})=0,\; k=0,1,2,\dots\label{eq:backward_Euler}
\end{equation}
where $I$ is an identity operator and $\tau_{k}=t_{k+1}-t_{k}$ is
a time step. Now,there are two main issues in solving the equation
\eqref{eq:backward_Euler} for $\hat{f}(x,y,t_{k+1})$. On each time
step we deal with: 
\begin{itemize}
\item \emph{Linearization}: find a linear equation that mimics the behaviour
of \eqref{eq:backward_Euler} locally with respect to $(x,y)$.
\item \emph{Approximation}: find an approximate solution that satisfies
the linear equation up to certain error tolerance. 
\end{itemize}
Let ${\mathcal{L}_{0}^{'}}_{\hat{f}}:W^{2}\rightarrow W$ be the Frechét
derivative \cite{Chang2005} of $\mathcal{L}_{0}$ at point $f(x,y,t_{k})$,
\begin{equation}
\begin{array}{rl}
{\mathcal{L}_{0}^{'}}_{\hat{f}}h= & K_t\bigg(T_{0,1}\,(f*xh+h*xf)-\mu_{x}f\cdot h-(\mu f+\lambda)x\cdot h\bigg)+{}\\
 & \rho K_t\bigg(T_{1,-1}\,(f*yh+h*yf)-\mu_{y}f\cdot h-(\mu f+\lambda)y\cdot h\bigg).\;\;\;{}
\end{array}
\end{equation}
Here, the direction $h\in W$ of Frechét derivative should not be
confused with the time step $\tau_{k}$. Now, a functional Newton's
method may be applied to \eqref{eq:backward_Euler} that defines a
sequence $\{\hat{f}^{s}\}_{0}^{\infty}$ converging to the root $\hat{f}(x,y,t_{k+1})$
of equation \eqref{eq:backward_Euler}, 
\begin{equation}
\hat{f}^{s+1}=\hat{f}^{s}-h_{s}.\label{eq:Newton}
\end{equation}
Here, we assume that the time step index $k$ is fixed and the initial
estimation $\hat{f}^{0}(x,y,t_{k+1})$ is given. Now, with each iteration
$s=0,1,2,\dots$, the new estimate $\hat{f}^{s+1}$ is obtained as
an upgrade of $\hat{f}^{s}$ with a certain correction term $h_{s}$
that solves the equation 
\begin{equation}
(\tau_{k}{\mathcal{L}_{0}^{'}}_{\hat{f}^{s}}-I)h_{s}=(\tau_{k}\mathcal{L}_{0}-I)\hat{f}^{s}+\hat{f}(x,y,t_{k}).\label{eq:Newton2}
\end{equation}
Although the equations (\ref{eq:Newton}, \ref{eq:Newton2}) contain
$\mathcal{L}_{0}$, they are linear with respect to the correction
term $h_{s}$. However, the estimated solution $\hat{f}^{s+1}$ is
expressed only implicitly as shown in \eqref{eq:Newton2}. Thus, we
proceed by approximately solving \eqref{eq:Newton2} employing a collocation
projection into a Gaussian basis\cite{Kryven2012}.

Let $a(h,v):W^{2}\rightarrow\mathbb{R},\: b(v):W\rightarrow\mathbb{R}$
be a bilinear and a linear form, respectively, being associated with
the right and left hand sides of \eqref{eq:Newton2}, 
\begin{equation}
\begin{array}{l}
a(h,v):=<(\tau_{k}{\mathcal{L}_{0}^{'}}_{f^{s}}-I)h,v>,\\
b(v):=<(\tau_{k}\mathcal{L}_{0}\hat{f}^{s}-I)+\hat{f}_{t_{k})},v>,
\end{array}\label{eq:forms}
\end{equation}
where $<\cdot,\cdot>$ is an inner product 
\begin{equation}
<f(x,y),g(x,y)>=\int\limits _{\Omega}f(x,y)g(x,y)\, dx\, dy.
\end{equation}
The equations (\ref{eq:Newton}, \ref{eq:Newton2}) may be reformulated
in a \emph{weak} sense 
\begin{equation}
\left\{ \begin{array}{l}
\hat{f}^{s+1}=\hat{f}^{s}-h_{s},\; h_{s}\in W;\\
a(h,v)=b(v),\;\forall v\in W.
\end{array}\right.\label{eq:weak_infinite}
\end{equation}
The idea behind a 'weak' formulation is the following: instead of
requiring the residual in the original equation \eqref{eq:Newton2}
to be zero, we rather demand a zero inner product of the residual
and all test functions $v\in W$. This is formulated in the second
line of \eqref{eq:weak_infinite}.

Finally, in order to perform the transition from the dimensionally
infinite $W$ to the finite degrees of freedom (DoF), we construct
a system of $n$ basis function centres $(x_{i},y_{i})\in\Omega$
and connectivity parameters $\sigma_{\cdot,i}$. Two $n$-dimensional
bases in $W$ are defined as follows 
\begin{equation}
\phi_{i}(x,y)=e^{-\sigma_{x,i}(x-x_{i})^{2}-\sigma_{y,i}(y-y_{i})^{2}},\;\sigma_{\cdot,i}>0,\; i=1,2,\dots,n.\label{eq:basis}
\end{equation}
\begin{equation}
\psi_{i}(x,y)=\delta(x-x_{i})\delta(y-y_{i}),\; i=1,2,\dots,n.\label{eq:test_basis}
\end{equation}
\begin{equation}
W_{n}=span\{\phi_{1},\phi_{2},\dots,\phi_{n}\}\,\subset W,\label{eq:W_n}
\end{equation}
\begin{equation}
W_{test}=span\{\psi_{1},\psi_{2},\dots,\psi_{n}\}\,\subset W,\label{eq:W_test}
\end{equation}
The basis functions \eqref{eq:basis} are used for expansion of the
approximation $\hat{h},\hat{f}^{s}\in W_{n},$where $\hat{h}$ approximates
$h$ and, to the weak solution $h\in W$,
\begin{equation}
\hat{h}(x,y)=\sum\limits _{i=1}^{n}\alpha_{i}\phi_{i}(x,y),\label{eq:anzats}
\end{equation}
\begin{equation}
\hat{f}^{s}(x,y)=\sum\limits _{i=1}^{n}\beta_{i}\phi_{i}(x,y),\label{eq:anzats2}
\end{equation}
while \eqref{eq:test_basis} is a source of test functions $v$ used
in \eqref{eq:weak_infinite}. Thus, we arrive at a discrete collocation
scheme for equation \eqref{eq:Newton2} 
\begin{equation}
a(h_{n},\psi_{j})=b(\psi_{j}),\; j=1,2,\dots,n\label{eq:Galerkin}
\end{equation}
For the sake of brevity we write $\boldsymbol{\alpha}=(\alpha_{i})$,
$\boldsymbol{\beta}=(\beta_{i})$ referring to the column vectors
of coefficients as defined in (\ref{eq:anzats}, \ref{eq:anzats})
and employ this matrix notation henceforth. Thus, for instance, relation
\eqref{eq:Galerkin} represents a system of $n$ linear equations
and by substituting \eqref{eq:anzats} and \eqref{eq:forms} into
\eqref{eq:Galerkin} one may bring it to the following matrix form
\begin{equation}
M^{0}\boldsymbol{\alpha}=\mathbf{b}\label{eq:linear_system}
\end{equation}
where $\boldsymbol{\alpha}$ is a column of unknown coefficients defining
the approximation to the correction term $\hat{h}^{s}$, while$M$
is a square $n\times n$ matrix, 
\begin{equation}
\begin{array}{ll}
M_{0} & =\tau_{k}K_t\bigg(\hat{T}_{0,1}\left(C_{\beta}\hat{T}_{x}+C_{\hat{T}_{x}\beta}\right)-\hat{\mu}_{x}\boldsymbol{\beta}I_{n}-(\hat{\mu}\boldsymbol{\beta}+\lambda)\hat{T}_{x}\bigg)+{}\\
 & {}+\tau_{k}\rho K_t\bigg(\hat{T}_{1,-1}\left(C_{\beta}\hat{T}_{y}+C_{\hat{T}_{y}\beta}\right)-\hat{\mu}_{y}\boldsymbol{\beta}I_{n}-(\hat{\mu}\boldsymbol{\beta}+\lambda)\hat{T}_{y}\bigg)-I_{n}.
\end{array}\label{eq:M}
\end{equation}
Here, $I_{n}$ is an identity matrix of size $n$, 
\begin{equation}
\hat{T}_{x_{s},y_{s}}=A^{-1}A_{x_{s},y_{s}}\label{eq:shift_operator}
\end{equation}
is a discrete approximation to the operator \eqref{eq:shift}, 
\begin{equation}
\begin{array}{lcl}
(A)_{i,j} & = & \phi_{j}(x_{i},y_{i})\\
(A_{x_{s},y_{s}})_{i,j} & = & \phi_{j}(x_{i}-x_{s},y_{i}-y_{s})
\end{array},\;\; i,j=1,2,\dots,n\label{eq:interpolation_operator}
\end{equation}
Matrices 
\begin{equation}
\begin{array}{l}
\hat{T}_{x}=A^{-1}\text{diag}\{x_{1},x_{2},\dots,x_{n}\}A,\\
\hat{T}_{y}=A^{-1}\text{diag}\{y_{1},y_{2},\dots,y_{n}\}A
\end{array}\label{eq:weight_operator}
\end{equation}
represent the multiplication with weights $x$ or $y$, respectively.
Matrices $C_{\beta},\, C_{\hat{T}_{w}\beta}$ represent the convolution
with the known approximation $\hat{f}^{s}$ or its weighted form $w\hat{f}^{s}$
associated to the $s$-th iteration of the Newton process \eqref{eq:Newton}
on the $k$-th time step. 
\begin{equation}
\begin{array}{rl}
C_{\beta}= & A^{-1}C,\\
(C)_{i,j}= & \sum\limits _{k=1}^{n}\beta_{k}\phi_{j}(x_{i},y_{i})*\phi_{k}(x_{i},y_{i}).
\end{array}\label{eq:discrete_convolution}
\end{equation}

Analogously to \eqref{eq:M}, by expanding \eqref{eq:forms} the column
vector at the right-hand-side of \eqref{eq:linear_system} may be
obtained as a matrix expression 
\begin{multline}
\mathbf{b}=\bigg(2\tau_{k}K_t\Big(\,\hat{T}_{0,1}C_{\beta}\hat{T}_{x}-\mu_{x}\boldsymbol{\beta}I_{n}-(\mu\boldsymbol{\beta}-\lambda)\hat{T}_{x}\,\Big)+{}\\
{}+\tau_{k}\rho K_t\Big(\,\hat{T}_{1,-1}C_{\beta}\hat{T}_{y}-\mu_{y}\boldsymbol{\beta}I_{n}-(\mu\boldsymbol{\beta}-\lambda)\hat{T}_{y}\,\Big)-I_{n}\bigg)\boldsymbol{\beta}\;+\boldsymbol{\beta}_{t_{k}},\label{eq:b}
\end{multline}
Finally, the partial moments $\mu_{x}^{k}(t),\,\mu_{y}^{k}(t)$ of
the approximated distribution may be obtained by the relation 
\begin{align}
(\boldsymbol{q})_{i}=\int_{\Omega}\phi_{i}(x,y)\, dx\, dy,\\
\begin{array}{ll}
\mu_{x}^{k}(t)= & \,\boldsymbol{q}^{T}T_{x}^{k}\boldsymbol{\beta}.\\
\mu_{y}^{k}(t)= & \,\boldsymbol{q}^{T}T_{y}^{k}\boldsymbol{\beta}.
\end{array}
\end{align}

The considerations expressed so far show how the approximation to
the distribution $f_{0}(x,y,t)$ of cyclized molecules may be retrieved
by solving the differential equation involving $\mathcal{L}_{0}$.
The situation regarding $f_{1}(x,y,t)$ is even simpler as the differential
equation \eqref{eq:f1} has a linear form, provided $f_{0}$ is known.
Thus, a linear matrix transform of the coefficient column vector $\boldsymbol{\alpha}_{t_{k}}$
on time step $t_{k}$ is sufficient to obtain data on time step $t_{k+1}$
. 
\begin{equation}
\boldsymbol{\alpha}_{t_{k+1}}=\boldsymbol{\alpha}_{t_{k}}+\tau_{k}M_{\beta_{k+1}}^{1}\boldsymbol{\alpha}_{t_{k+1}}
\end{equation}

\begin{equation}
M_{\beta}^{1}=2K_t(C_{\beta}\hat{T}_{x}-\boldsymbol{q}^{T}\boldsymbol{\beta}\hat{T}_{x})+\rho K_t(C_{\beta}\hat{T}_{y}-\boldsymbol{q}^{T}\boldsymbol{\beta}\hat{T}_{y})+\lambda K_t(T_{x}\boldsymbol{\beta}+\rho T_{y}\boldsymbol{\beta})
\end{equation}

Until now, only a first order time integration has been considered.
The numerical tools derived in the previous section may provide sufficient
data for higher order time integration techniques. For instance, the
computational codes belonging to multistep methods for stiff systems
require subroutines computing the coefficients vector for the discretized
right hand side of \eqref{eq:theeq}, $\boldsymbol{\alpha}_{c}=\hat{\mathcal{L}_{c}}\boldsymbol{\beta},\;\hat{\mathcal{L}_{c}}:\mathbb{R}^{n}\rightarrow R^{n}$,
and its Jacobian matrix $J_{\hat{\mathcal{L}_{c}}}\boldsymbol{\beta}$.
Here, we strongly benefit from the specific construction of the test
function space (\ref{eq:test_basis},~\ref{eq:W_test}) that satisfies
\[
<f(x,y),\psi_{j}(x,y)>=f(x_{j},y_{j}),\;\psi_{j}\in W_{test}.
\]
Furthermore, the explicit formulas for the coefficients are obtained
analogously to the expressions (\ref{eq:M},~\ref{eq:b}),

\begin{equation}
\begin{array}{rl}
\hat{\mathcal{L}_{0}}\boldsymbol{\beta}= & \bigg(2K_t\Big(\,\hat{T}_{0,1}C_{\boldsymbol{\beta}}\hat{T}_{x}-\mu_{x}I_{n}-\mu\hat{T}_{x}\,\Big)+{}\\
 & {}+K_t^{'}\Big(\,\hat{T}_{1,-1}C_{\boldsymbol{\beta}}\hat{T}_{y}-\mu_{y}I_{n}-\mu\hat{T}_{y}\,\Big)\bigg)\boldsymbol{\beta};\\
J_{\hat{\mathcal{L}_{0}}}\boldsymbol{\beta}= & 2K_t\bigg(\hat{T}_{0,1}\left(C_{\boldsymbol{\beta}}\hat{T}_{x}+C_{\hat{T}_{x}\boldsymbol{\beta}}\right)-\mu_{x}I_{n}-\mu\hat{T}_{x}\bigg)+{}\\
 & {}+K_t^{'}\bigg(\hat{T}_{1,-1}\left(C_{\boldsymbol{\beta}}\hat{T}_{y}+C_{\hat{T}_{y}\boldsymbol{\beta}}\right)-\mu_{y}I_{n}-\mu\hat{T}_{y}\bigg).
\end{array}\label{eq:beta}
\end{equation}
In the case of $\mathcal{L}_{1}$ the corresponding differential equation
\eqref{eq:f1} is linear, hence 
\begin{equation}
\hat{\mathcal{L}}_{1}\boldsymbol{\beta}=J_{\hat{\mathcal{L}}_{0}}\boldsymbol{\beta}=M_{\beta}^{1}.
\end{equation}
In conclusion, at this stage we have developed a fully discrete set
of formulas that may be implemented into computational code in a straightforward
manner or treated employing existing stiff numerical integrators (e.g.
\emph{MATLAB ode13s}).

\subsection{The numerical algorithm}

In order to provide a summary of the instructions listed in the previous
section, the numerical algorithm is depicted in the diagram of Appendix
1. The discretization techniques presented fully describe the approximated
systems by a collection of column vectors $\boldsymbol{\beta}_{t_{k}},\: k=0,1,2,\dots$
that correspond to each time step $t_{k}$. Regarding space approximations,
the basis function centres are enumerated with a single integer number,
so notwithstanding the dimensionality of two, the problem could be
parametrized using a vector of $n$ coefficients at a single time
point. Furthermore, Newton's iteration process is established at every
time step $t_{k}$, so we introduce an upper index $s$ to refer to
the sequence of estimations in the Newton's process: $\boldsymbol{\beta}_{t_{k}}^{s}$.

\begin{figure}
\begin{centering}
\includegraphics[width=0.5\textwidth]{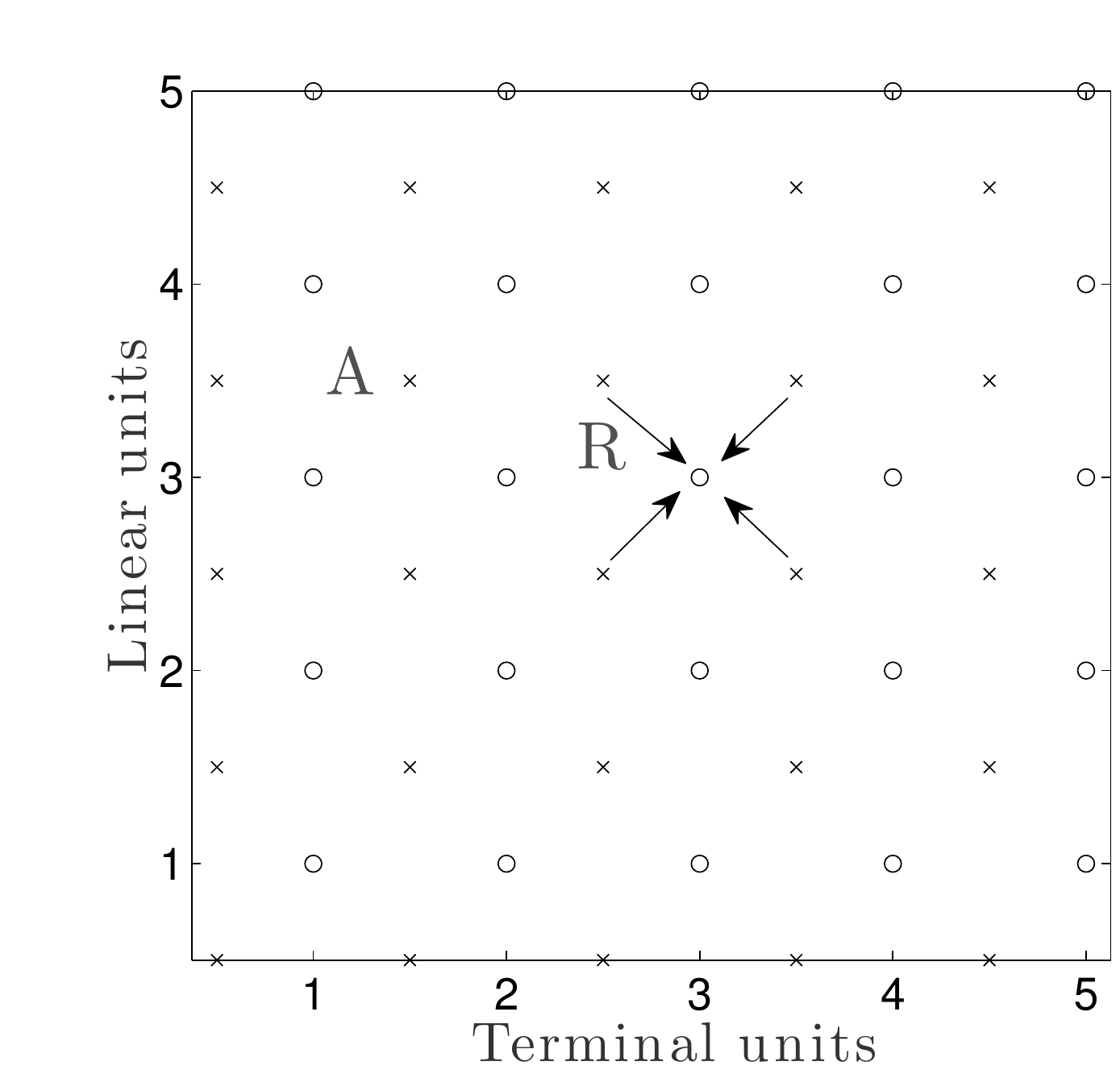} 
\par\end{centering}

\caption{The principle of adaptive mesh refining is based on an error evaluation
at intermediate nodes \textsf{x}. The original nodes system is denoted
by \textsf{o}}

\label{fig:AdaptiveMesh} 
\end{figure}

Before the algorithm starts, one must set the basis parameters $x_{i},\, y_{i},\,\sigma_{x,y}$.
This could be realized either on the basis of the previous simulations
in order to refine the results in certain parts of the domain, or
as a logarithmically distributed system that covers the whole domain
of possible values $(x,y)$, $[0,x_{max}]\times[1,y_{max}]$. There
is no simple way to compute optimal values for $\sigma_{x,y}$, although
the parameters should be dependent on the distance between two adjustment
basis functions. Since the collocation approach tends to give the
smallest absolute error at the basis function centres, the approximation
error could be a posteriori evaluated locally, by repeating the simulations
using additional checkpoint basis functions in between the original
ones. This principle is illustrated in Figure~\ref{fig:AdaptiveMesh}.
The the approach of residual subsampling\cite{driscoll2007} is applied
to refine the mesh. 
\begin{itemize}
\item If the local error is greater than a certain tolerance $Tol_{add}$,
the intermediate checkpoint node $A$ is accepted in the new system. 
\item By default all original basis centres are accepted in the new system. 
\item If the local error in four intermediate nodes is smaller than a predefined
tolerance $Tol_{remove}$, the original node $R$ situated in between
is not accepted into the new system. 
\end{itemize}

The matrix expression derived in the paper could become computationally
intensive. Nevertheless, as it is shown in \eqref{eq:M} and \eqref{eq:b},
the matrices $M$ and $b$, dependent on time and a certain intermediate
result, are to be recomputed on every step and consist of constant
numbers (\ref{eq:shift_operator}, \ref{eq:interpolation_operator},
\ref{eq:weight_operator}). that only depend on the fixed basis functions
parameters, and hence should only be computed once at the initialization
of the algorithm.

\section{Results and post-analysis}

\begin{figure}
\center \includegraphics[width=0.7\textwidth]{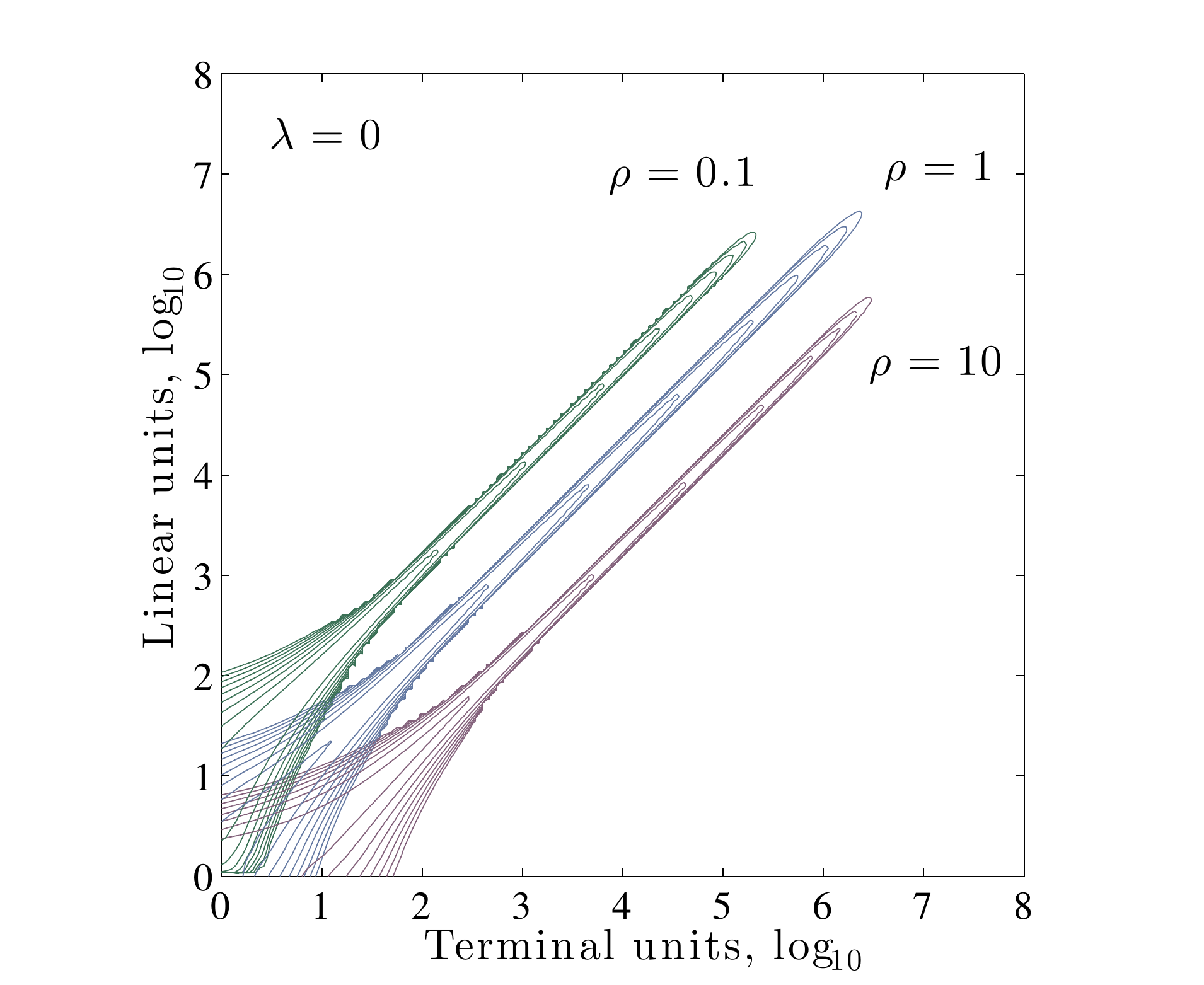} \caption{The level-lines of weighted frequency distribution $xyf_{0}(x,y,t_{end})$
as obtained from simulations. Different values for the substitution
factor are considered $\rho\in\{0.1,1,10\}$. For each case 8 level-lines
are plotted $10^{-(4.5+0.5k)},k=1,\dots,8$.}

\label{fig:2d_fxy_rho} 
\end{figure}

\begin{figure}
\center \includegraphics[width=0.6\textwidth]{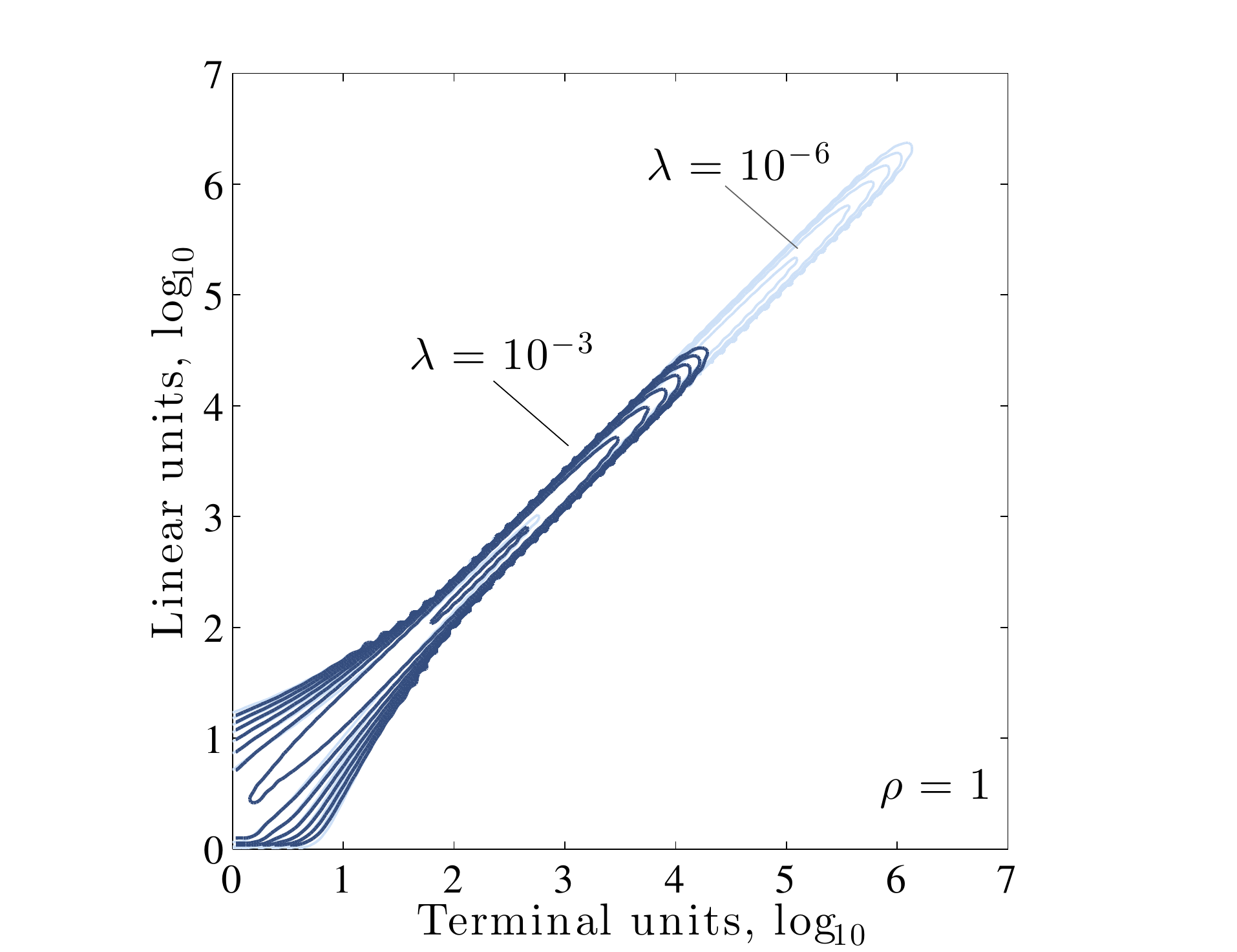} \caption{The level-lines of weighted frequency distribution $xyf_{0}(x,y,t_{end})$
as obtained from simulations. Different values for the cyclization
parameter are considered $\lambda\in\{10^{-3},10^{-4},10^{-6}\}$.
In each case 10 level-lines are plotted $10^{-(4.2+\frac{1}{2}k)},k=1,\dots,8$.}

\label{fig:2d_fxy_lambda} 
\end{figure}
The numerical scheme described in the previous paragraph shows many advantages when comparing with existing simulation techniques. It provides a representation of the full distribution, that is accurate enough to perform data mining even in the regions with very small values. Despite high precision, the method remains computationally inexpensive. As a result we are able to extract very detailed morphology related properties by post-processing. For instance information on cycle length, that was typically associated with Monte Carlo simulations before, is obtained in a fully deterministic manner.

\begin{figure}
\center \includegraphics[width=0.6\textwidth]{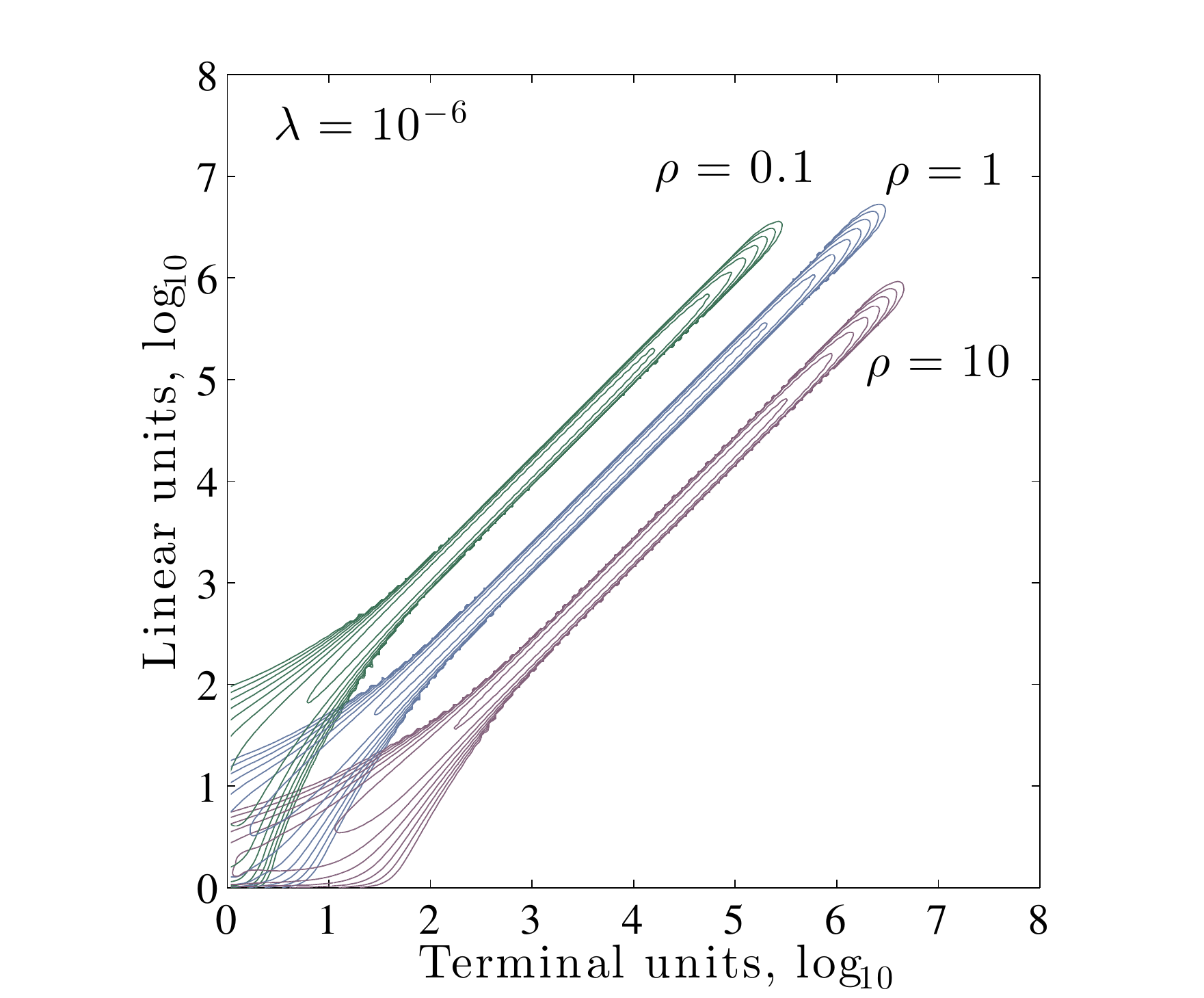} \caption{The level-lines of cyclized molecules frequency weighted distribution
$xyf_{1}(x,y,t_{end})$ as retrieved from $f_{0}(x,y,t)$. Different
values for the substitution factor are considered $\rho\in\{0.1,1,10\}$.
In each case 7 level-lines are plotted $10^{-(5+\frac{1}{2}k)},k=1,\dots,10$.}

\label{fig:2d_fxyc_rho} 
\end{figure}

By following the numerical scheme we generate full time profiles for non-cyclized and cyclized molecules
$f_{0}(x,y,t),f_{1}(x,y,t)$ that correspond to the model \eqref{eq:theeq}
up to a conversion of A groups $c=99.95\%$. The full set of parameters is given in Table~\ref{tab:parameters}.   
The effect of two input
parameters is studied in detail: 
\begin{itemize}
\item $\rho$~the ratio between the reactivity of substituted and non-substituted
groups, see Figure~\ref{fig:2d_fxy_rho};
\item $\lambda$~the ratio of the cyclization rate to the polymerization
rate, see Figure~\ref{fig:2d_fxy_lambda}. 
\end{itemize}
The distribution for cyclized structures $f_{1}(x,y,t)$ is retrieved
by integrating \eqref{eq:f1}, Figure~\ref{fig:2d_fxyc_rho}. 

\begin{table}
\begin{centering}
\begin{tabular}{l|l}
$\rho$ & 0.1, 1, 10\tabularnewline
$\lambda$ & $0,$ $10^{-6},$ $10^{-5},$ $10^{-4},$ $10^{-3}$ \tabularnewline
Conversion of A groups, $c$  & 0.9, 0.99, 0.997, 0.998, 0.999, 0.9999 \tabularnewline
Degree of freedom for approximation & 1200 \tabularnewline
$x_{i},y_{i}$ & $\{1,2,3,2^{k},k=2\dots8log_{2}10\}$ \tabularnewline
$\sigma_{x,i}$  & $\{2.7(x_{i}-x_{i-1})^{-2},i=1,\dots,n\}$ \tabularnewline
Tolerance for Newton iterations $\epsilon$ & $10^{-14}$ \tabularnewline
Initial time step & $10^{-10}$ \tabularnewline
computational time, single core CPU:  & \tabularnewline
\hspace{0.5cm}for quadratures (to be done once) & 35~min.\tabularnewline
\hspace{0.5cm}for time integration (to be done for each parameter
setup) & 5~min.\tabularnewline
\end{tabular}
\par\end{centering}

\caption{Parameters used in simulations}

\label{tab:parameters} 
\end{table}

The results shown in Figures~\ref{fig:2d_fxy_rho},\ref{fig:2d_fxyc_rho}
already provide an idea how the substitution factor $\rho$ influences
the topologies. Indeed, varying the values for $\rho$ is observed
shifting the two dimensional distribution diagonally along the diagonal
line $\log x+\log y=\text{const}$. This equally holds for non-cyclized
molecules Figure~\ref{fig:2d_fxy_rho}, and cyclized molecules Figure~\ref{fig:2d_fxyc_rho}.
The cyclization factor $\lambda$ shifts the distribution along $\log x-\log y=\text{const}$
as shown in Figure~\ref{fig:2d_fxy_lambda}. This - at first sight
uncomplicated - behaviour of the two dimensional distribution nevertheless
has dramatic implications for the scalar and distributive properties
that are to be inferred by post processing.

In view of high amount of statistical properties that are extractable
from the simulation results $f_{c}$, we group them into four categories:
time dependent scalars and 1,2,3-dimensional distributions.

\subsection{Time dependent scalars}

\begin{figure}[H]
\center \includegraphics[width=0.5\textwidth]{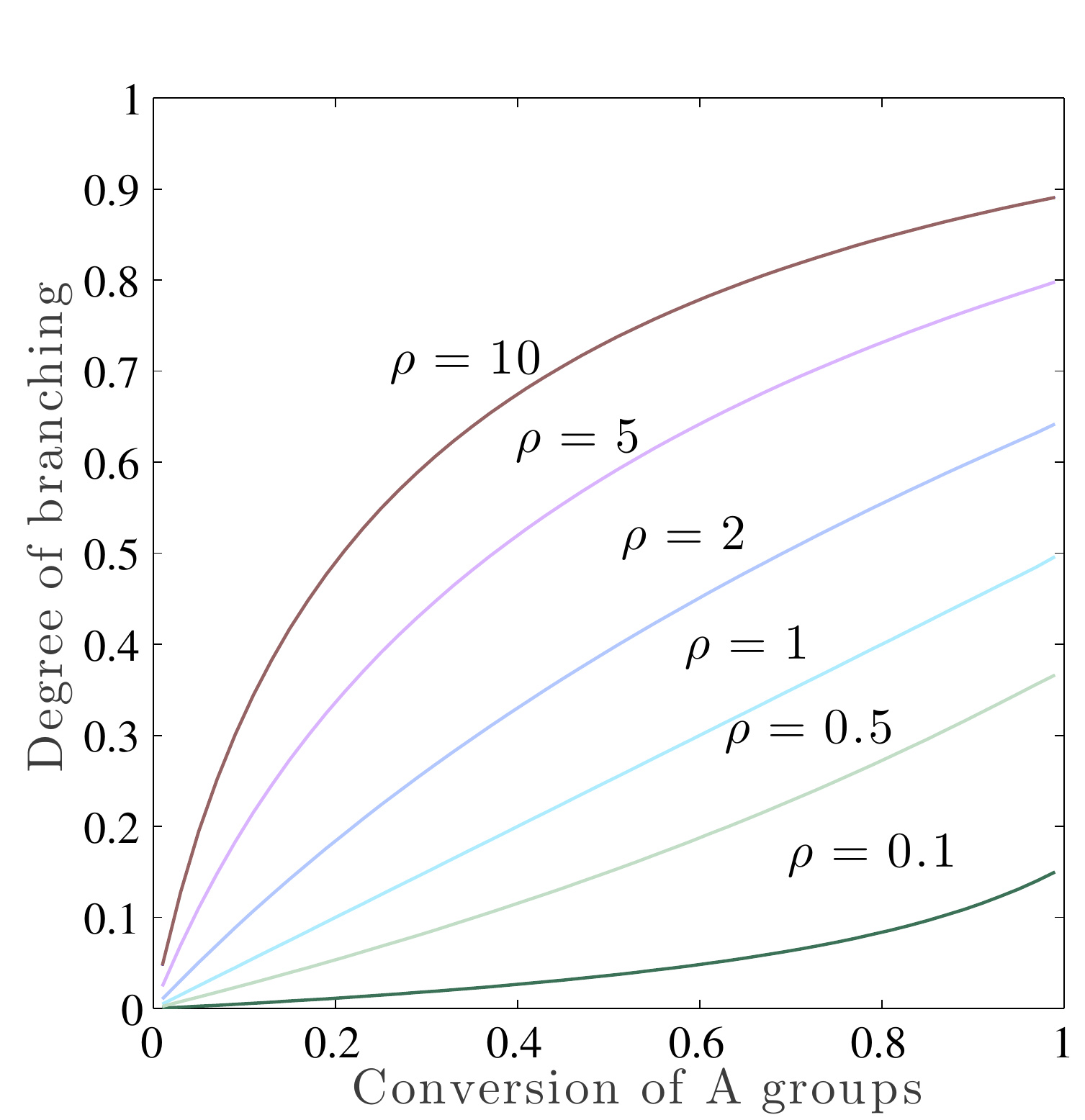}
\caption{Degree of branching $db(t)$ as a function of conversion of A groups
$c(t)$ in polymerization of $AB_{2}$ monomers reacting with a substitution
effect for B-groups$\rho$.}

\label{fig:scalar_db} 
\end{figure}

\begin{figure}[h]
\center \includegraphics[width=0.6\textwidth]{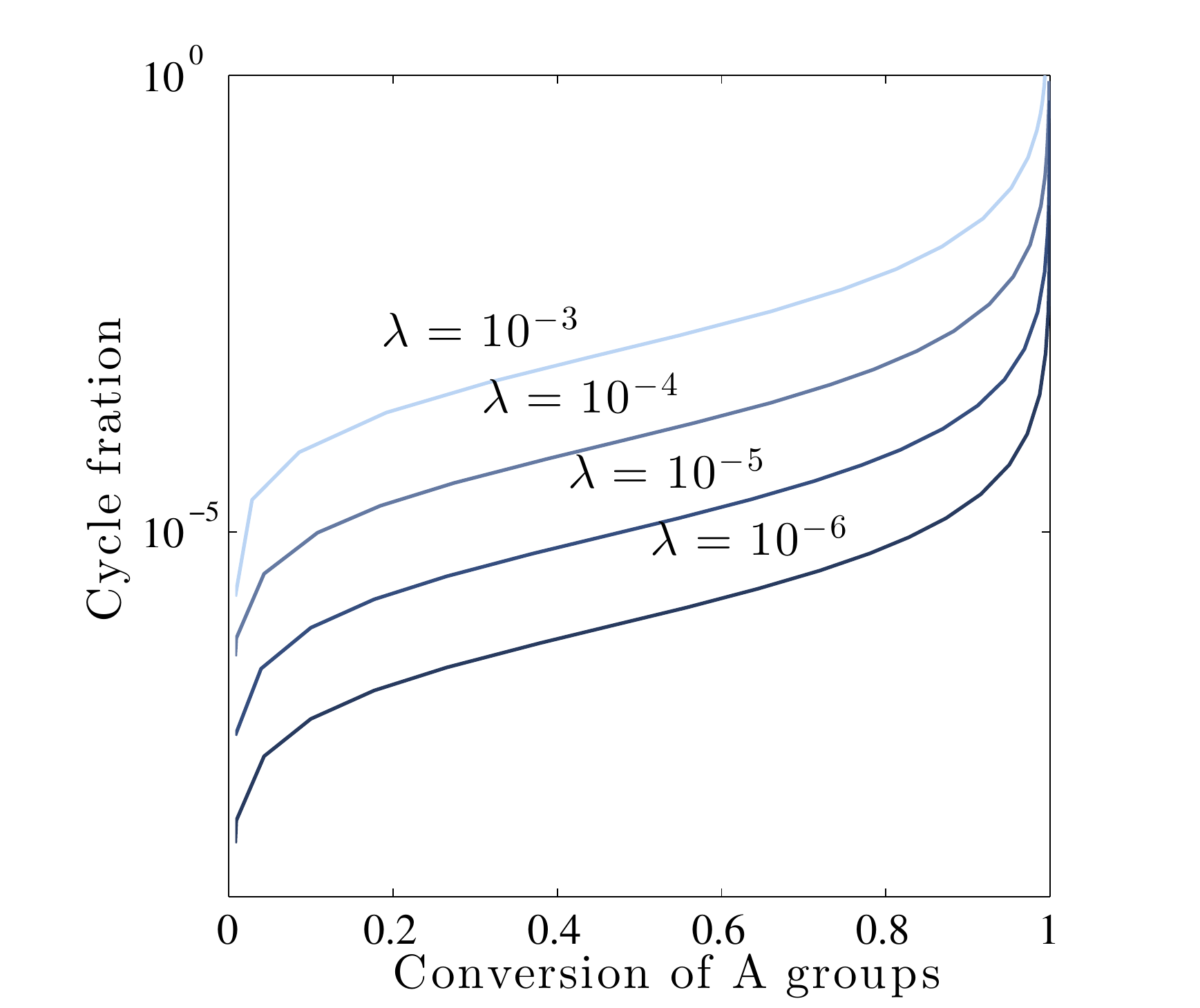} \caption{Fractions of cycle-containing molecules $n_{C}(t)$ versus conversion
$c(t)$ as calculated from the simulation. The parameter $\lambda$
controls the extent of cyclization. The substitution effect $\rho$
does not affect this property.}

\label{fig:cycle_fraction} 
\end{figure}

The time dependent scalars are obtained from the moments \eqref{eq:moments}
of the density $f_{c}(x,y,t)$ as a convenient consequence of basic
graph properties \eqref{eq:trivial}. All time dependent values will
be presented with respect to conversion $c\in[0,1)$ of the free $A$
groups, so $K_t$ coefficient may be chosen arbitrary. Indeed, the
time dynamics for the conversion of A groups $c(t)$; the number of
free A groups $n_{A}(t)$; the number of linear $n_{L}(t)$, terminal
$n_{T}(t)$ or dendritic $n_{D}(t)$ units; the fraction of cycles
$n_{C}(t)$ or the degree of branching may be expressed as follows:
\begin{equation}
\begin{array}{rl}
c(t) & =\mu\big(f_{0}(x,y,t)+f_{1}(x,y,t)\big);\\
n_{A}(t) & =1-c(t);\\
n_{L}(t) & =\mu_{y}f_{0}(x,y,t);\\
n_{T}(t) & =\mu_{x}f_{0}(x,y,t);\\
n_{D}(t) & =n_{T}(t)-n_{A}(t);\\
n_{C}(t) & ={\mu f_{1}(x,y,t)}/{c(t)};\\
db(t) & =\frac{n_{D}}{n_{D}+0.5n_{L}}.
\end{array}\label{eq:scalars}
\end{equation}
Here we illustrate the most important of the listed properties. The
Frey's degree of branching\cite{Frey1997} $db(t)$ turns out being
strongly affected by the substitution factor $\rho$, while the cyclization
parameter $\lambda$ does not influence the results significantly,
Figure~\ref{fig:scalar_db}. On the other hand, the fraction of cyclized
molecules $n_{C}(t)$ is dependant on $\lambda$ but not on $\rho$,
Figure~\ref{fig:cycle_fraction}. This results are in good agreement
with previous findings\cite{Galina2002,Cheng2002}. 

\subsection{1-dimensional distributions}

\begin{figure}[H]
\center 
\begin{tabular}{rc}
a) & \includegraphics[clip,width=0.5\textwidth]{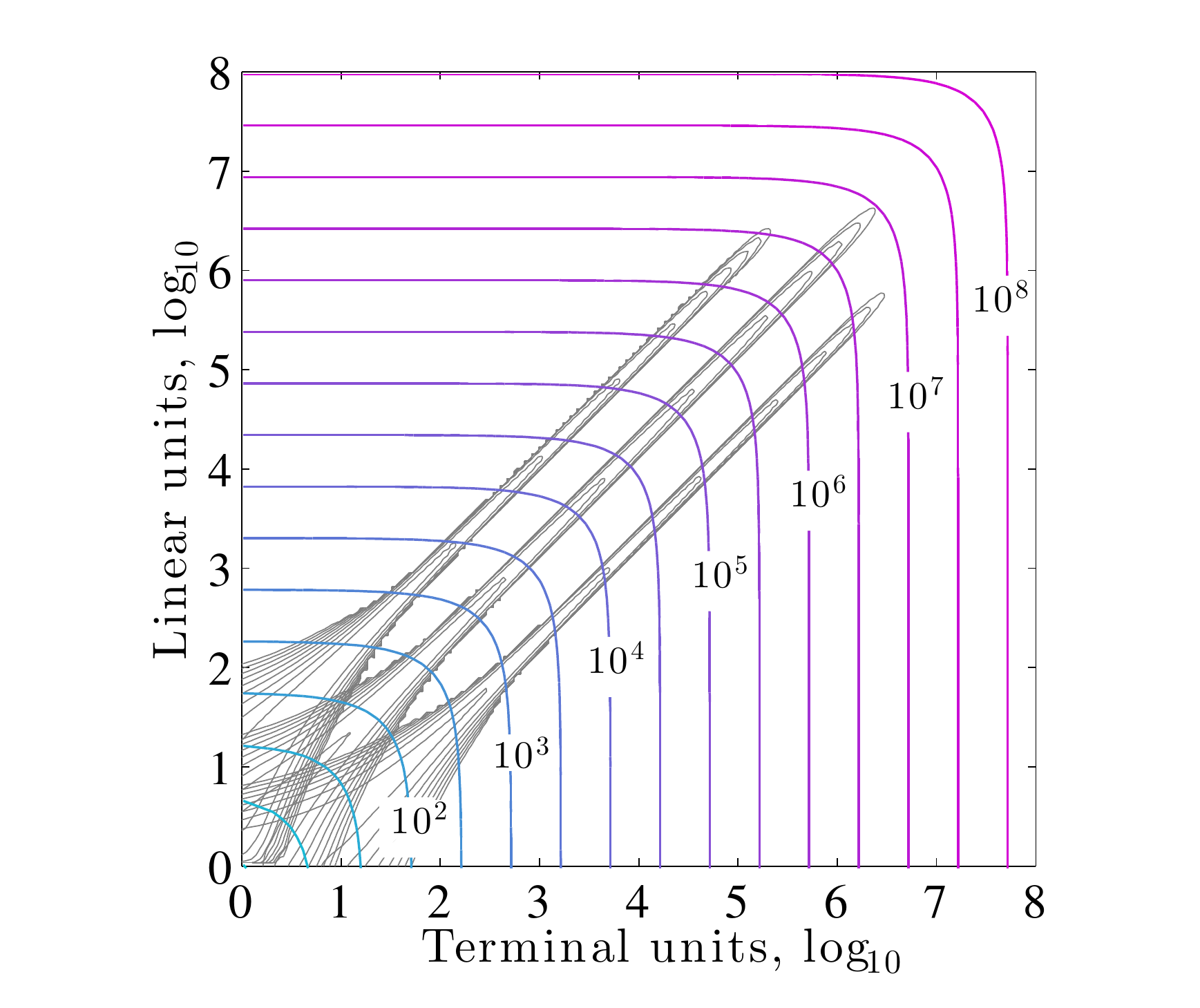} \\
b) & \includegraphics[width=0.5\textwidth]{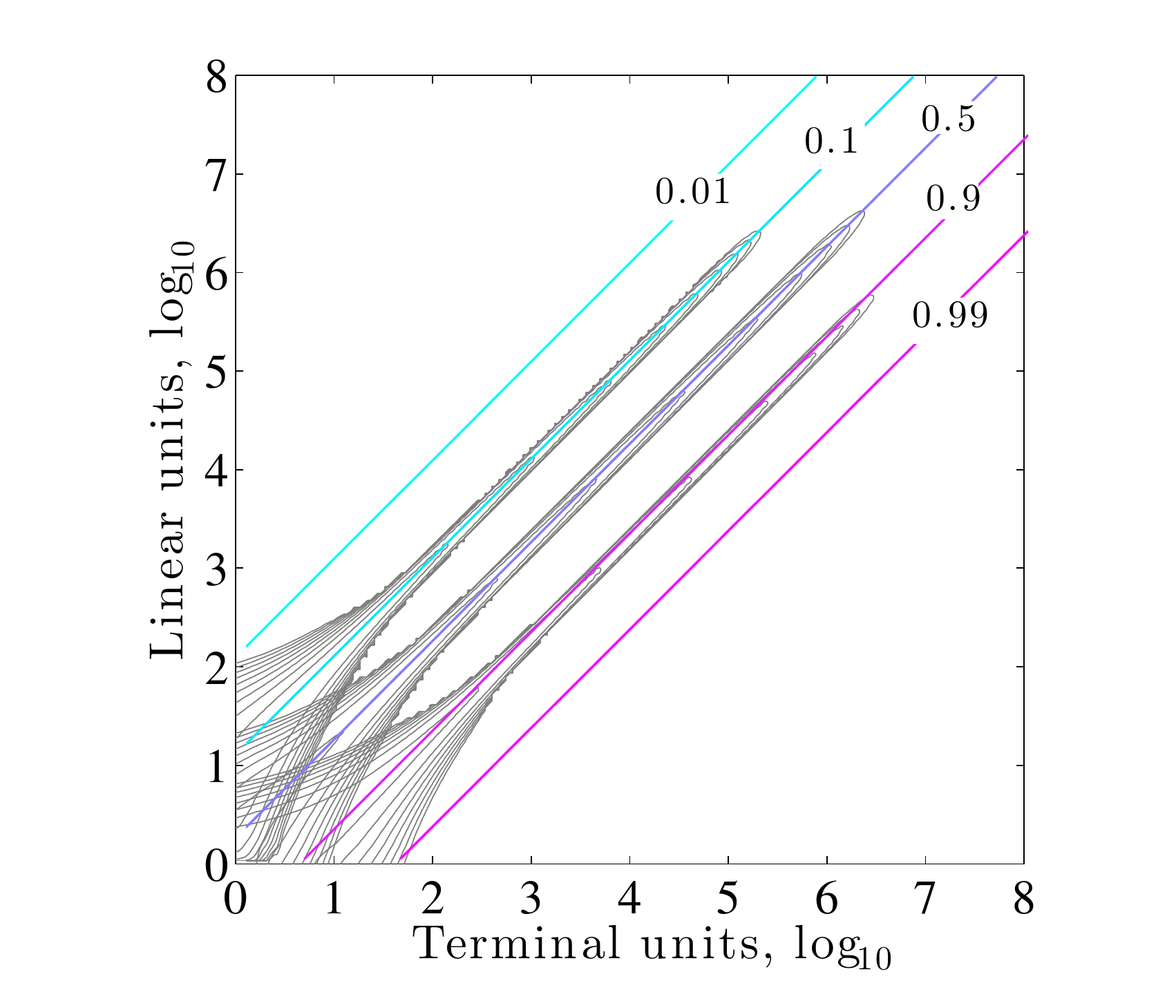} \\ 
c) & \includegraphics[width=0.5\textwidth]{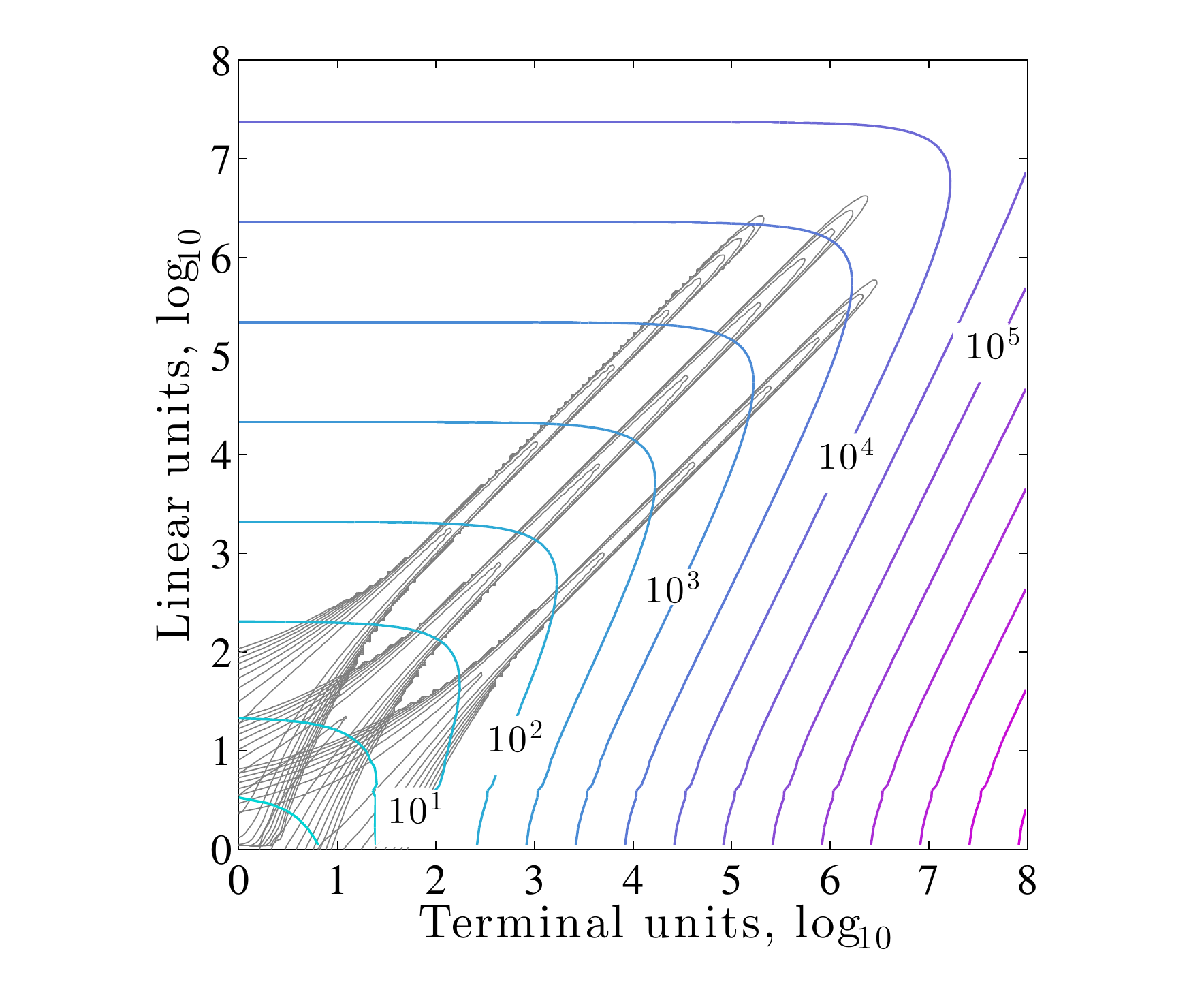}
\end{tabular}
\caption{Three collections of lines that keep constant values of a) chain length
, b) degree of branching, c) length of a cycle. The level lines of
weighted frequency distribution $xyf_{0}(x,y,t_{end})$ with $\lambda=0,\,\rho=\{0.1,\,1,\,10$
are plotted in the background for reference.}

\label{fig:lines} 
\end{figure}

One-dimensional distributions are obtained by evaluation of line integrals
of $f_{c}(x,y,t_{end})$ for a fixed time point $t_{end}$. Distributions
of molecular weight, degree of branching, or cycle lengths may be
obtained in this fashion by defining a collection of lines that keeps
a certain property constant, see Figure~\ref{fig:lines}. 

\begin{figure}[h]
\center \includegraphics[width=1\textwidth]{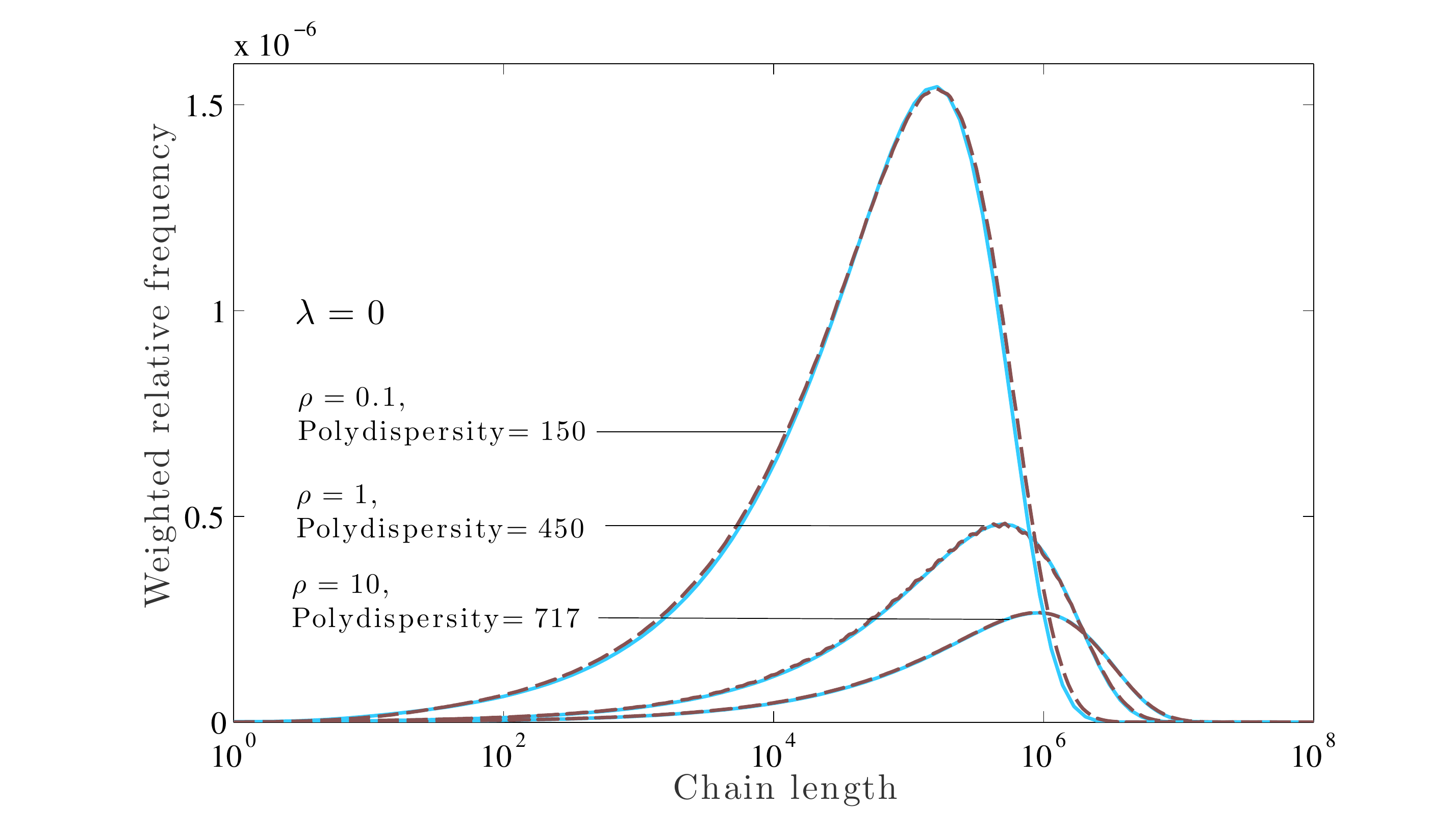} \caption{Normalised double weighted chain length distribution $n^{2}ld(n)$.
The predicted values are indicated by the dashed lines and compared
to the analytical solution, the solid lines. The effect of three levels
of chemical substitution $\rho\in\{0.1,1,10\}$ is illustrated under
the assumption that no cyclization takes place.}

\label{fig:MWDs} 
\end{figure}

\begin{figure}[h]
\center \includegraphics[width=0.6\textwidth]{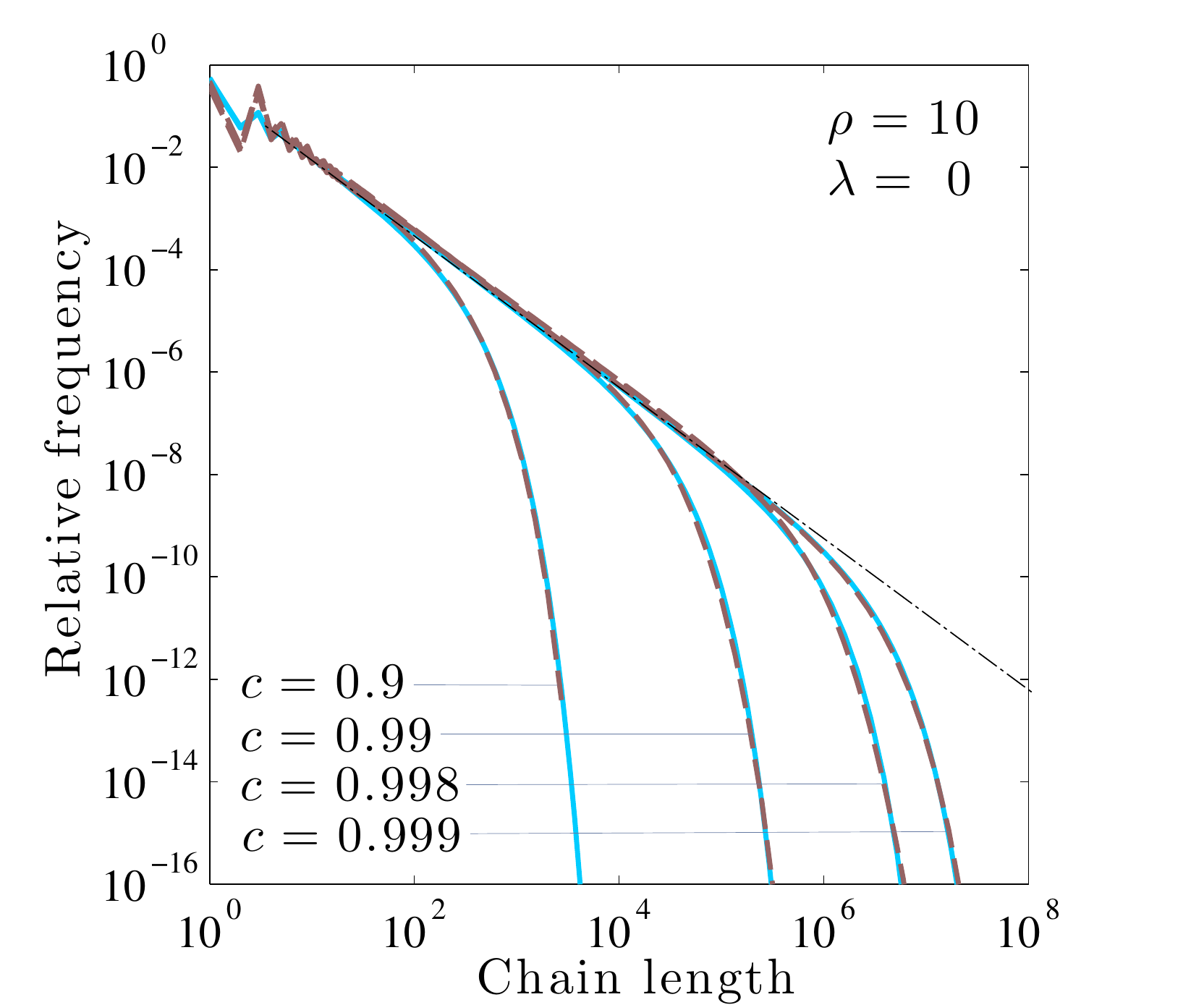} \caption{Chain length distribution $ld(n)$ as predicted by the numerical method,
indicated by the solid lines, as compared to the analytical solution,
in dashed lines. The effect of different values for conversion $c\in\{0.9,\,0.99,\,0.998,\,0.999\}$
is illustrated under the assumption that no cyclization takes place
$\lambda=0$. The asymptotic value for conversion 1 is illustrated
by the dash-dot line. }

\label{fig:MWDsLog} 
\end{figure}

For instance, in order to recover the amount of molecules with a certain
length $n$ from $f_{0}(x,y,t_{end})$, one has to consider all combinations
of points $(x,y)$ that satisfy $2x+y-1=n$ (see Figure~\ref{fig:lines}a),
and sum the corresponding values of $f_{0}$. Mathematically, this
is equivalent to evaluation of a line integral along the collection
of lines $ld_{n}(y)=0.5(n-y+1),\, y\in[0,n-1]$, 
\begin{equation}
ld(n)=\frac{1}{2}\int\limits _{0}^{n-1}f_{0}(ld_{n}(y),y,t_{end})\, dy\label{eq:length_distribution}
\end{equation}
The molecular weight distribution $n^{2}ld(n)$ for various substitution
factors and no cyclization is shown in Figure~\ref{fig:MWDs}. The
effect of different conversion values as the distribution is approaching
its asymptotic values is shown in Figure~\ref{fig:MWDsLog}. The
effect of approaching a straight line in a double logarithmic plot
as a system approach the gelation point was predicted before for a
simpler systems\cite{wattis2006}. It is also important to consider
the cases without cyclization, as an analytical solution is known
\cite{Zhou2006}, which may be employed to validate the numerically
obtained results. As shown in Figures~\ref{fig:MWDs},\ref{fig:MWDsLog}
perfect agreement is observed over a span of 16 orders of magnitude.

\begin{figure}[H]
\center \includegraphics[width=0.7\textwidth]{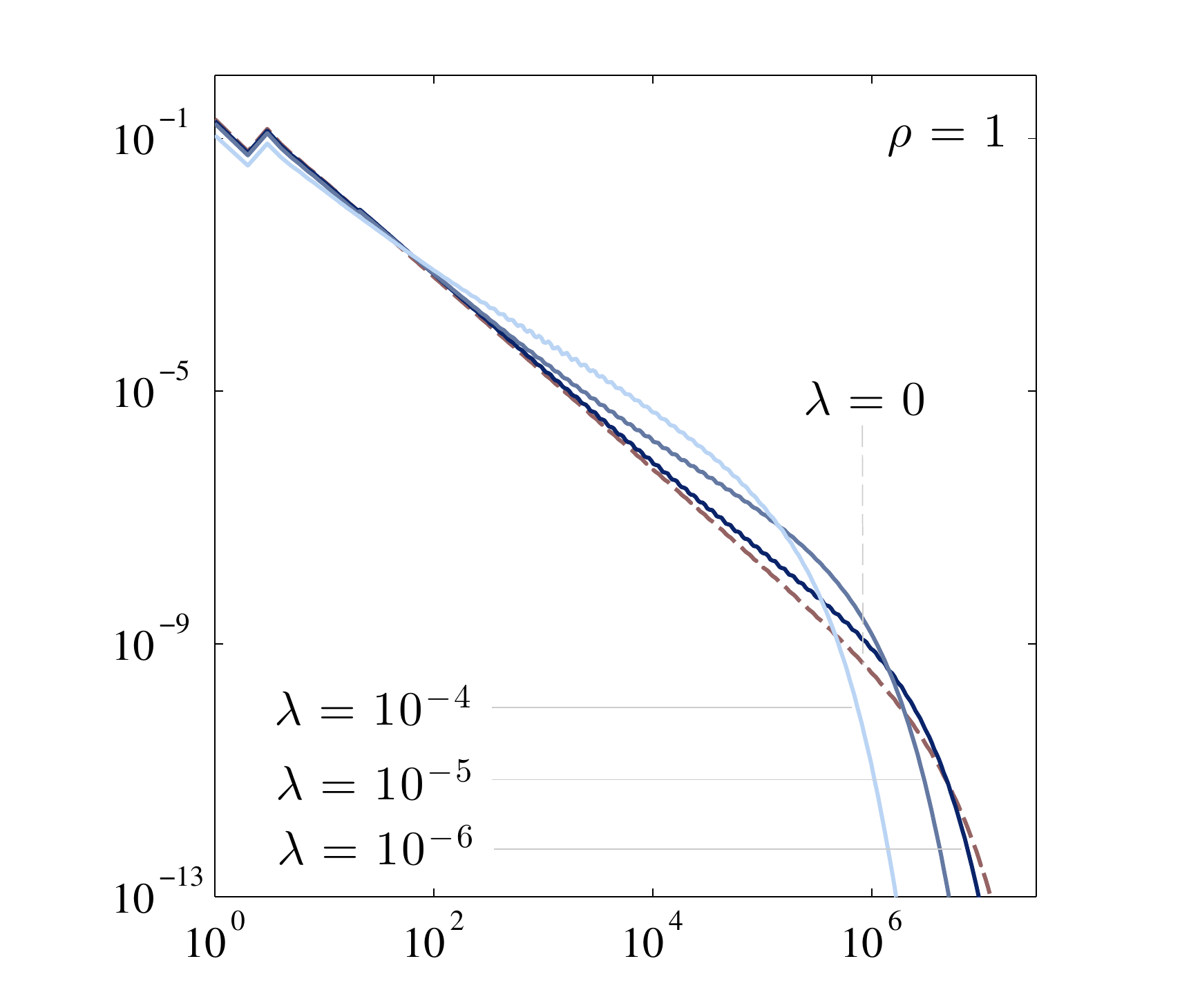} \caption{The effect of the different rates of cycle formation $\lambda\in\{10^{-6},10^{-5},10^{-4}\}$
on the chain length distribution. The dashed line depicts the case
without cycle formation. }

\label{fig:MWD_lambda} 
\end{figure}

The effect of the cyclization factor $\lambda$ on the chain length
distribution for a fixed conversion value and substitution factor
is shown in Figure~\ref{fig:MWD_lambda}.

\begin{figure}[h]
\center \includegraphics[width=0.6\textwidth]{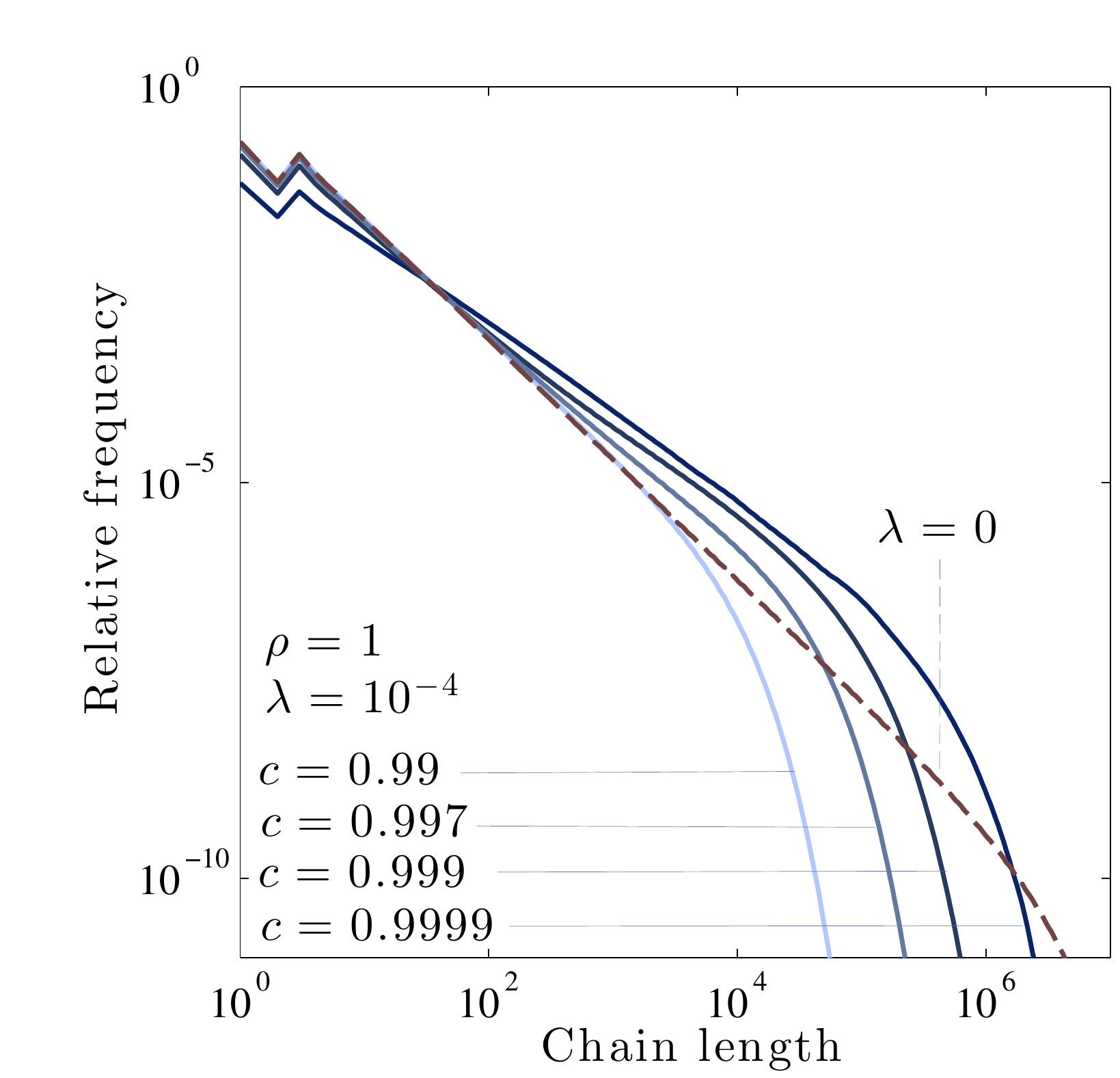} \caption{Chain length distribution $ld(n)$ as predicted by the numerical method.
The solid lines denote cases with a cyclization factor $\lambda=10^{-4}$,
for different values of conversion. The dashed line depicts the non-cyclization
case for reference. The effect of different values of conversion $c\in\{0.99,\,0.997,\,0.999,\,0.9999\}$
is illustrated. }

\label{fig:MWD_cycled} 
\end{figure}

\begin{figure}[h]
\center \includegraphics[width=0.7\textwidth]{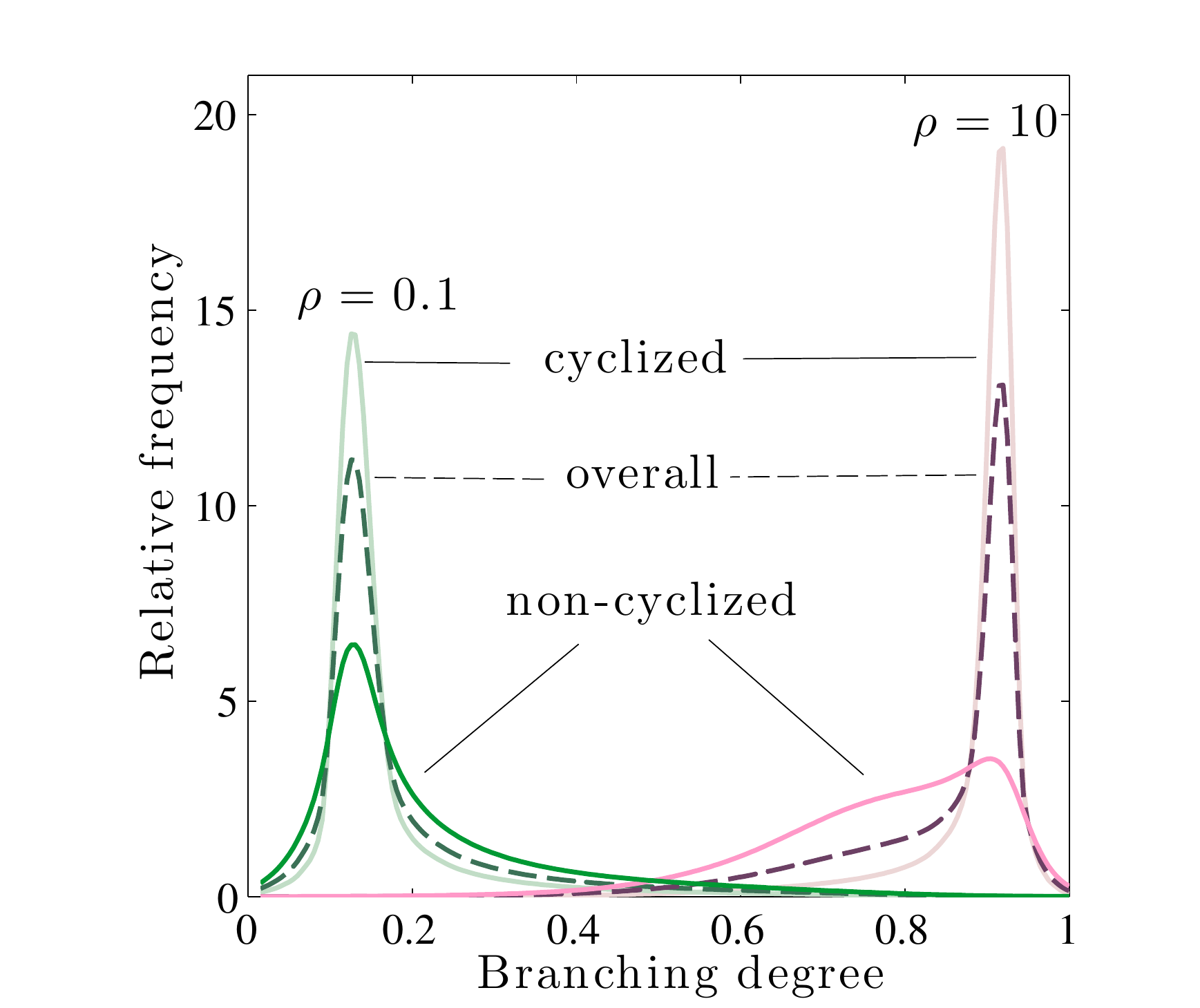} \caption{Normalised Frey's degree of branching distribution as obtained from
the simulations for two levels of chemical substitution $\rho\in\{0.1,10\}$.
The overall distribution (the dashed lines) is plotted together with
its two components, the distributions for cyclized and non-cyclized
polymers (solid lines). The rate of cycle formation is $\lambda=10^{-4}$. }

\label{fig:db} 
\end{figure}

The length distributions of polymers with different degree of branching
$b\in(0,1]$ is found by integrating over the collection of lines
defined by $l_{b}(x)=2\frac{1-b}{b}x,\, x\in[1,\infty)$ (see Figure~\ref{fig:lines}b)
s Analogously to \eqref{eq:length_distribution} we obtain 
\begin{equation}
bd(b)=\int\limits _{0}^{\infty}(2x+y)f_{c}(x,l_{b}(x),t_{end})\, dx.
\end{equation}
Note, we intentionally use the chain length $2x+y$ as a weight while
integrating in order to magnify the contribution of longer molecules,
that have much more monodisperse branching distribution. Although,
the weight was used we still observe prolonged tails in the branching
distribution $bd(b)$ Figure~\ref{fig:db}. This fact has to be accounted
for when one uses average an value instead of a distribution to describe
the branched topology of the $AB_{2}$ system. We also observe that
the cyclized molecules contribute less to the tails than non-cyclized
molecules. 

\begin{figure}[h]
\center \includegraphics[width=0.6\textwidth]{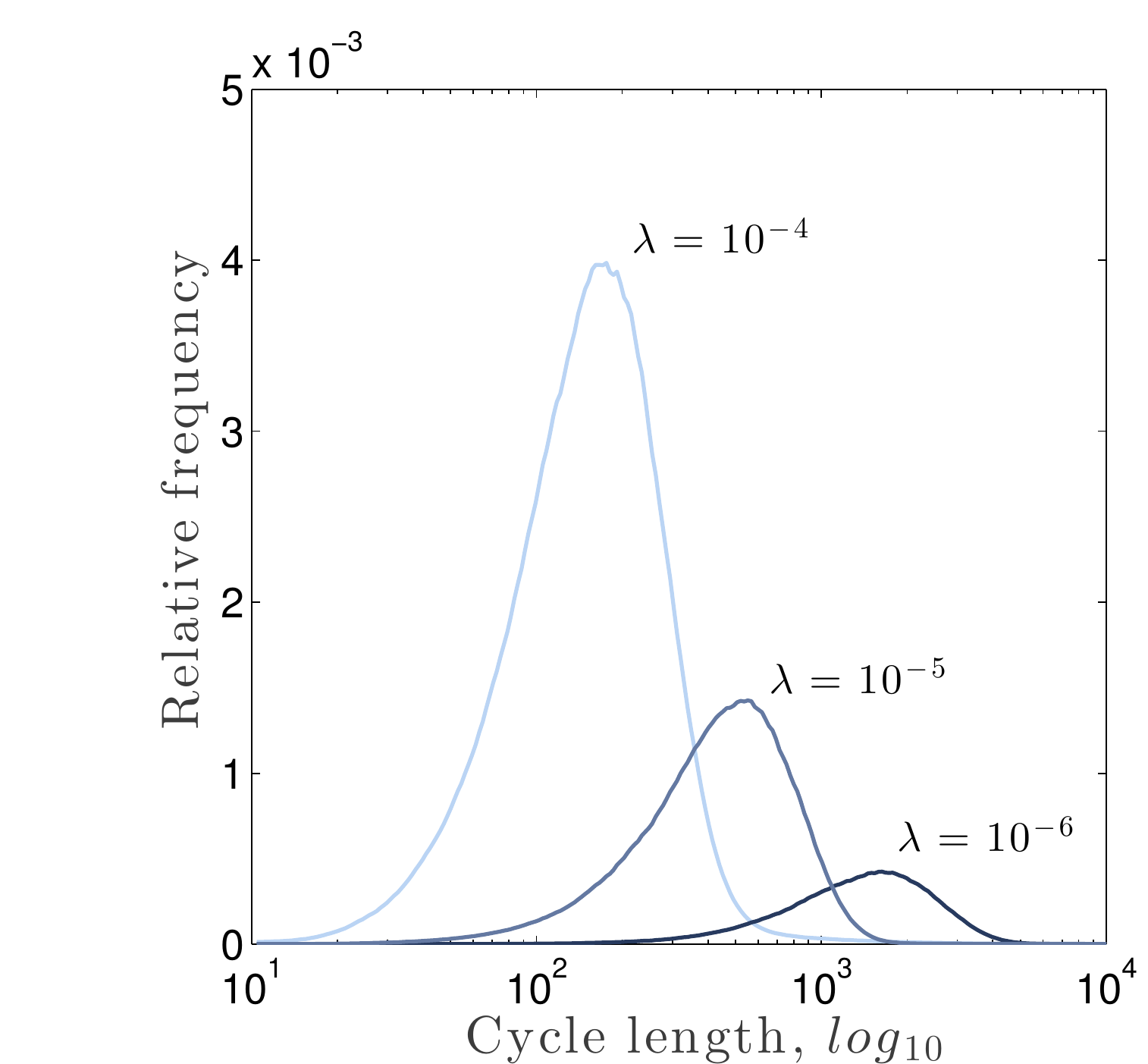} \caption{Normalised cycle-length distribution as obtained from the integration
of $f_{c}(x,y,t_{end})$. The effect of different values for cyclization
factor $\lambda\in\{10^{-6},10^{-5},10^{-4}\}$ is illustrated. }

\label{fig:cycles} 
\end{figure}

In order to obtain information concerning \emph{cycle lengths, }we
employ an idea formulated by to Panholzer et al.\cite{panholzer2002}.
It was shown that the expectation of a distance $depth_{i,n}$ from
the root to an $i$-labelled terminal unit is statistically connected
to the number of terminal units $x$ in a \emph{strictly binary} tree:
\begin{equation}
depth_{i,x}=\frac{2(2i+1)(2x-2i+1)}{x+2}\frac{\binom{2i}{i}\binom{2x-2i}{x-i}}{\binom{2x}{x}}-1,\; i=0,\dots x.
\end{equation}
Now, by proportionally extending the length of each path by a number
of linear units and averaging over the label index $i$, an expected
cycle length $cl(x,y,c_{type})$ is obtained, 
\begin{equation}
cl(x,y,c_{type})=2^{-c_{type}}\frac{x+0.5y-1}{x^{2}-x}\sum\limits _{i=0}^{x}depth_{i,x}.\label{eq:cycle_lines}
\end{equation}
The level lines of \eqref{eq:cycle_lines} are illustrated in Figure~\ref{fig:lines}c.
Note that it is important to know whether the reaction that caused
cyclization added an edge between the root and a \emph{terminal} unit
$c_{type}=0$ or between the root and a \emph{linear} unit $c_{type}=1$.
As follows from the model \eqref{eq:operator_L0}, at each instant
of time $\lambda(xf_{0}+\rho yf_{0})$ new cyclized molecules appear.
The instantaneous cycle length distribution of these molecules may
be extracted according to principle \eqref{eq:cycle_lines}. In order
to retrieve the total accumulated distribution of cycle lengths we
have to additionally integrate over the whole time span $[0,t_{e}nd]$.
This requires the evaluation of the double integral, 
\begin{multline}
cd(n)=\int\limits _{0}^{t_{end}}\int\limits _{cl(x,y,0)=n}\lambda xf_{0}(cl(x,y),\tau)\, d(x,y)\, d\tau+\\
\int\limits _{0}^{t_{end}}\int\limits _{cl(x,y,1)=n}\lambda\rho yf_{0}(cl(x,y),\tau)\, d(x,y)\, d\tau.\label{eq:cycle_distribution}
\end{multline}
As in previous cases we evaluate the results \eqref{eq:cycle_distribution}
in a numerical manner. The result is depicted in Figure~\ref{fig:cycles}.
That it turns out to be possible to retrieve the cycle length distribution
proves the great potential of this deterministic method, especially
when taking the few dimensions considered into account. It provides
sufficient information to obtain morphology related properties that
are usually only attainable with Monte Carlo simulations.

\subsection{2-dimensional distributions}

\begin{figure}[H]
\center %
\begin{tabular}{ccc}
a)&\includegraphics[width=0.4\textwidth]{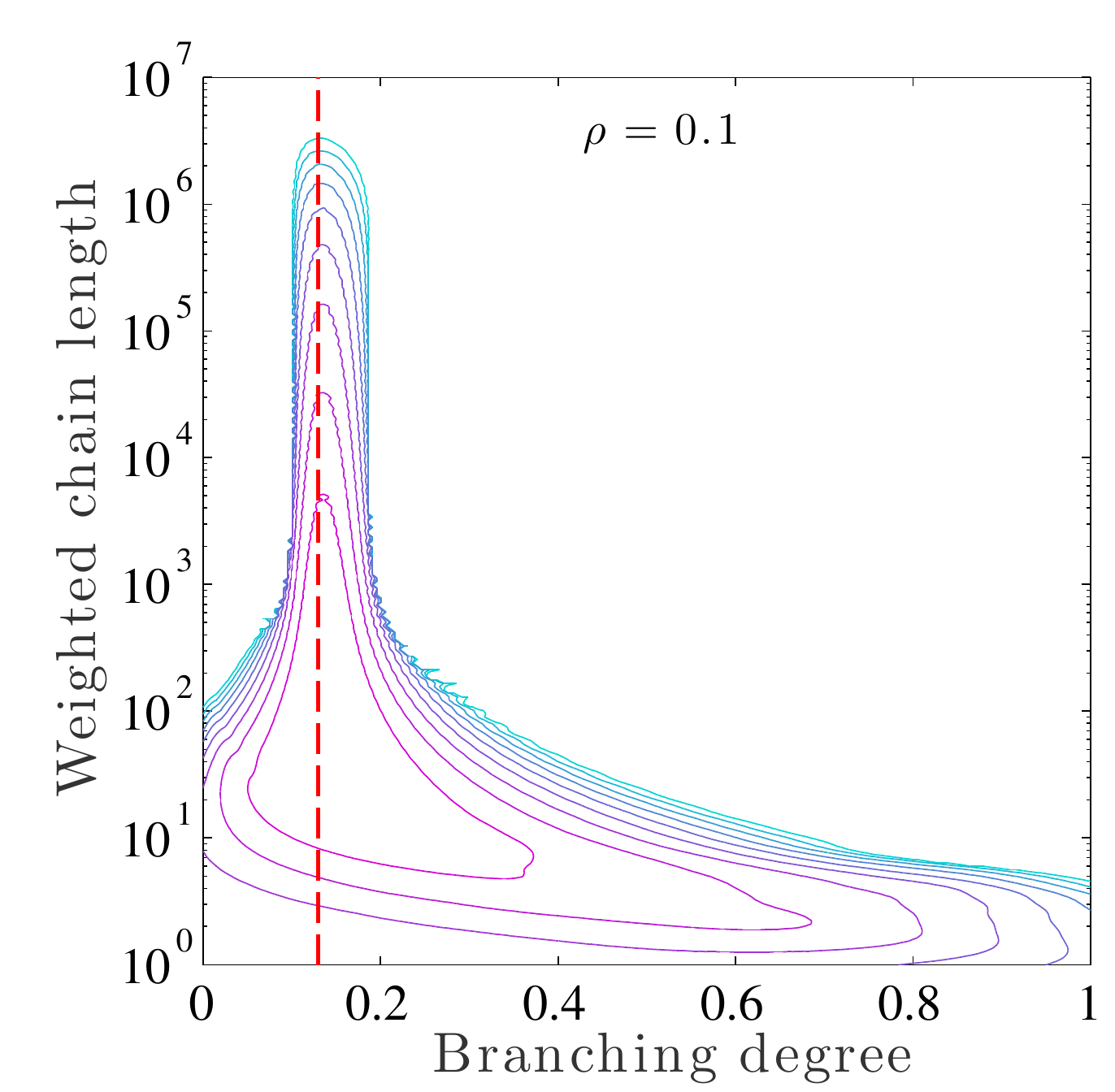}  \\
b)&\includegraphics[width=0.4\textwidth]{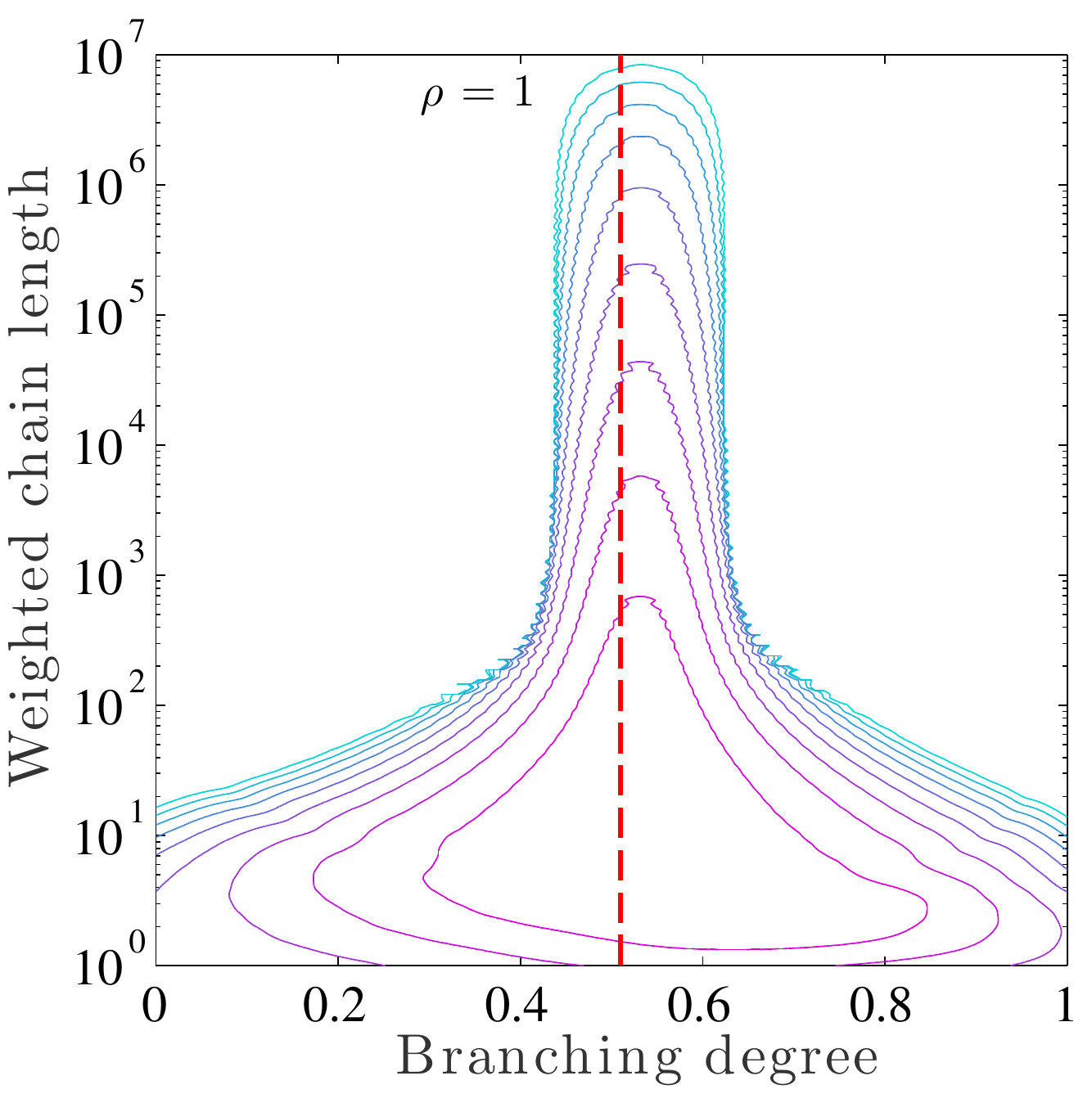} \\ 
c)&\includegraphics[width=0.4\textwidth]{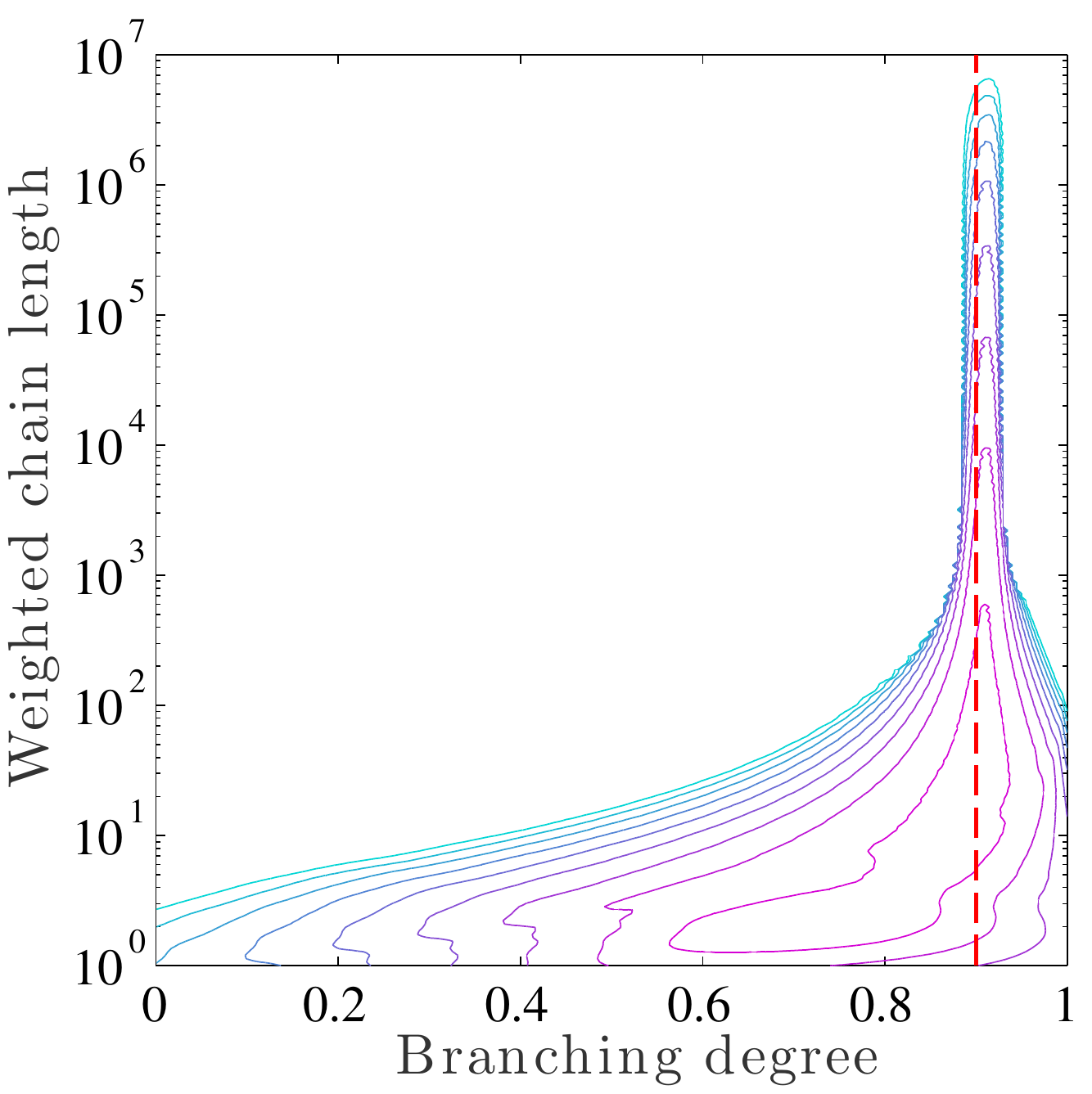}
\end{tabular}\caption{Two-dimensional distributions of chain length and degree of branching.
Three values for the substitution factor are considered, $\rho\in\{0.1,1,10\}$.
In each case 8 level lines are plotted, $10^{-(3+\frac{1}{2}k)},k=1,\dots,8$.
The dotted line depicts a scalar degree of branching for the whole
system.}

\label{fig:db_vs_length} 
\end{figure}

The distribution $f_{c}(x,y,t_{end})$ denotes the relative frequency
of structures with $x$ terminal units, $y$ linear units, and $c\in\{0,1\}$,
as depicted in Figures~\ref{fig:2d_fxy_rho},\ref{fig:2d_fxyc_rho}.
It is interesting to change the coordinate system of terminal units
versus linear units $(x,y)$ to another system. For instance, the
coordinate system chain length versus degree of branching $(d,n)$
may provide information that would remain unnoticed otherwise. In
this case the following transformation of coordinates is required,
\begin{equation}
\begin{array}{rcl}
x & \rightarrow & \frac{x}{x+0.5y}\\
y & \rightarrow & 2x+y-1
\end{array}
\end{equation}
According to \eqref{eq:scalars} the first line of the coordinates
transform replaces $x$ with Frey's degree of branching, while the
second line replaces $y$ with chain length. The resulting distribution
is depicted in Figure~\ref{fig:db_vs_length}. One may observe that
the distributions are centred around the average property: Frey's
degree of branching. However, they remain fairly broad for chain lengths
smaller than $10^{3}$, while accompanied by extremely narrow peaks
for longer chain lengths. This fact remains hidden when only the average
degree branching is considered.

\subsection{Dynamic evolution of 2D distributions}

\begin{figure}[H]
\center \includegraphics[width=0.8\textwidth]{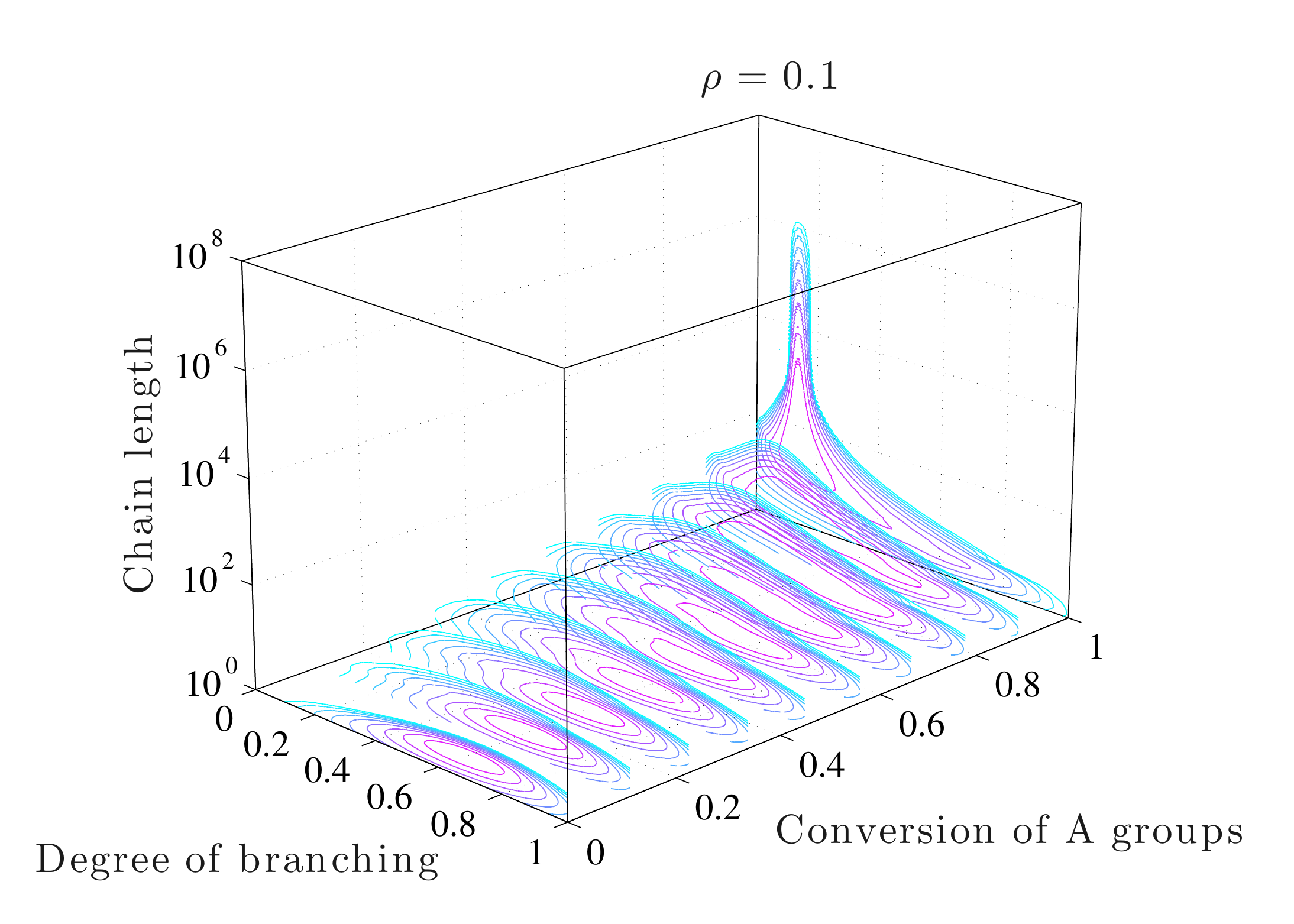} \includegraphics[width=0.8\textwidth]{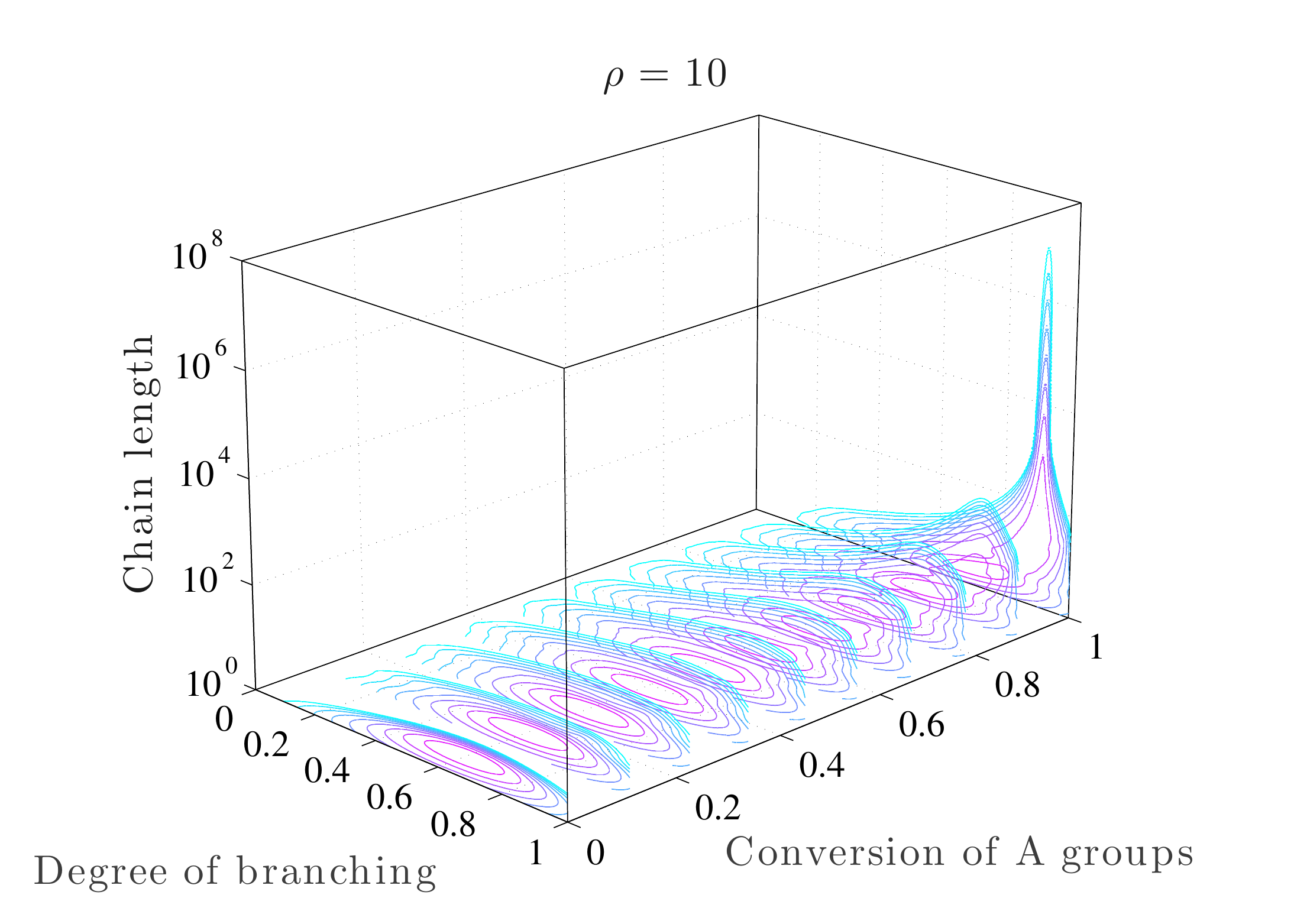}
\caption{Molecular structures with different degree of branches and molecular
weight distribution at 10 time-slices. Two values for the substitution
factor are considered, $\rho\in\{0.1,10\}$. In each case 10 level
lines are plotted, $10^{-(2+0.5k)},k=1,\dots,10$.}

\label{fig:3d} 
\end{figure}

As an example of 3-dimensional distribution obtained from the simulations,
we present a time trajectory for distributions shown in Figure~\ref{fig:db_vs_length}.
Here $f_{c}(x,y,t)$ is not only considered at the end time point
$t_{end}$ but similar to \eqref{eq:scalars} on the whole time interval
$t\in[0,t_{end}]$ as depicted in Figure~\ref{fig:3d}. Note, as
in the previous subsection we use double-weighted chain length in
order to highlight the long molecules present at extremely low concentrations.

\section{Non-linear substitution as a consequence of shielding }

As has been pointed out by Šomvarsky et. al. \cite{Somvarsky1998,Somvarsky1994}
the reactivity of a functional group may be dependant on the position
of the group within the molecule's topology due to a steric excluded
volume effect. The amount of non-shielded terminal $x'$ and linear
$y'$ units may be defined by employing a power law 
\begin{equation}
\begin{array}{l}
x'\propto x^{\omega};\\
y'\propto y^{\omega},
\end{array}
\end{equation}
where $\omega$ is the constant that describes the shape of the molecules,
as a measure for the reactive surface having a fractal character \cite{Somvarsky1998}.
The qualitative investigation of the kernels with fractional and integer
values for $\omega$ has been performed by Wattis \cite{wattis2006}.
The exactly solvable cases correspond only to integer values for $\omega$
\cite{wattis2006}. In case of a fractional $\omega$, the method
of generating functions\cite{Costa2007} would be particularly difficult
to apply as the transformation of the kernel into a domain of generating
functions would imply a \emph{fractional} differentiation, whose definition
usually requires an integral transformation, which in this case would
return the original (unsolved) problem. In the previous sections,
we have already considered the case of possibly substituted, but equally
accessible groups $\omega=1$ and as we shall see further on that
there are no particular difficulties to expand the approach to cases
of fractional $\omega$.

In the current paragraph the opposite marginal case of spheric particles
$\omega=\frac{2}{3}$ is studied. The numerical scheme allows accounting
for this shielding effect by applying only minor alterations to the
original equations. It suffices reformulating the weight operators
\eqref{eq:weight_operator} in the following manner 
\begin{equation}
\begin{array}{l}
\hat{T}_{x}^{\omega}=A^{-1}\text{diag}\{x_{1}^{\omega},x_{2}^{\omega},\dots,x_{n}^{\omega}\}A,\\
\hat{T}_{y}^{\omega}=A^{-1}\text{diag}\{y_{1}^{\omega},y_{2}^{\omega},\dots,y_{n}^{\omega}\}A
\end{array}\label{eq:omega_weight_operator}
\end{equation}

\begin{figure}[H]
\center \includegraphics[width=0.6\textwidth]{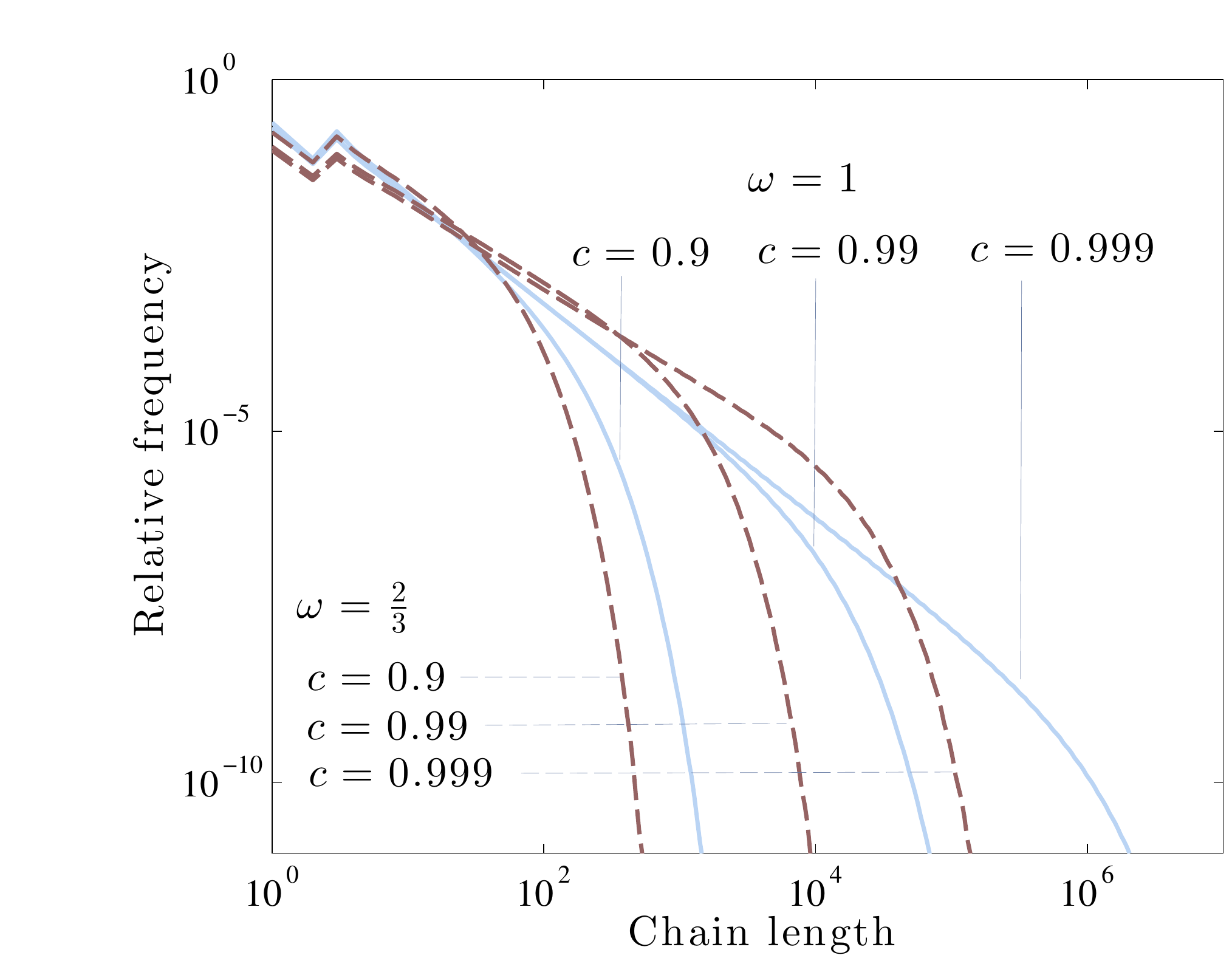} \caption{The chain length distributions at different levels of conversion,
$c\in{0.9,\,0.99,\,0.999}$ as obtained by numerical simulation accounting
for the shielding effect, $\omega=2/3$ (depicted with dashed line).
The chain length distributions for identical parameters, but without
shielding $\omega=1$ are depicted with solid lines for reference. }

\label{fig:MWD_omega} 
\end{figure}

By taking analogous steps as before we arrive at an approximate solution
to the polymerization problem with shielded groups. The chain length
distribution at different levels of conversion is presented in Figure~\ref{fig:MWD_omega}.
As a result, assuming non equal accessibility for reactive functional
groups leads to a slowing down of reaction rates, and consequently
shorter molecules are formed.

\section*{Conclusions}
The novel numerical method capable of capturing two-dimensional population balances rigorously, was developed as a result of clear mathematical reasoning. In contrary to prior results, the method reveals a full two-dimensional distribution. 
The method represents distributions as a linear combination of Gaussian basis functions, and finds their expansion coefficients, finally resulting in an approximate solution of the population balance equation. The accuracy of more than 15 orders-of-magnitude was observed when comparing with the analytical results.

The distributive properties of hyperbranched $AB_{2}$ system with
a substitution has been obtained by applying the method to a population balance model. In
contrast to prior modelling studies of $AB_{2}$ polymerization, the
distributive properties are obtained in multiple dimensions and maintain high resolution, giving access
to information of large molecules that are only present in extremely
low concentrations. In addition, an intramolecular cyclization reaction
has been accounted for in the model. The influence of the chemical
substitution and shielding as well as various cyclization rates on
the molecular topologies has been studied.

The numerical results obtained for the case of no cyclization turned
out to be in excellent agreement with the analytical solution present
in the literature\cite{Zhou2006}. For the cyclization case we observe
agreement with averaged properties obtained by Galina\cite{Galina2002}.
However, the newly developed method is able to reveal much more detailed
information. Interestingly, the simulations show that the degree of
branching is strongly dependent on the substitution factor, while
the chain length is predominately determined by cyclization rate.
Low cyclization rates cause small amounts of molecules with long cycle
length and vice versa, high cyclization causes more molecules to have
cycles, but shorter in length. It was also found that the branching
distribution has a broad non-trivial shape that is non-symmetrical
for cases with substitution effect. The distribution can be relatively
well described by Frey's degree of branching index (FDB) for long
molecules where it deviates less from the expected value. In contrast,
for short molecules the deviation increases and FDB only poorly describes
topologies. It has also been observed that the degree of branching
for molecules with cycles is generally more narrowly distributed than
for non-cyclized molecules.

Cyclization turns out to have two different effects on the chain length
distribution. Higher rates of cyclization tend to shorten the tail
at very long chain length, but at the same time the chain length distribution
decays more slowly in the mid range. The effect of the cyclization
on chain length and branching distributions may be explained by the
fact that cyclized molecules react at a lower rate, which attributes
a kind of 'memory' to the polymerization system and blurs the distributions
towards the earlier time stages.

It was also shown how the chemical substitution implemented as a linear
factor, can be adapted to model the shielding effect. The shielding
effect is implemented by a non-linear kernel with a fractional power,
and assumes only those functional groups are accessible for reaction
that are situated on the surface with dimension of fractal character. 

The developed algorithm provides a fast and efficient way for predicting
multidimensional distributive properties for hyperbranched polymers.
This was successfully demonstrated on the example of $AB_{2}$ type
polymerization accounting for the substitution effect and cyclization.
Although the paper is more than just a study of a particular molecular
system, we have presented a methodology that is capable of handling
a wide spectrum of polymer related problems and is especially helpful
in cases of more than one dimension and convolution equations.

\bibliographystyle{elsarticle-num.bst}
\bibliography{literature_AB}

\section{Appendix 1}

\includegraphics[width=0.9\textwidth]{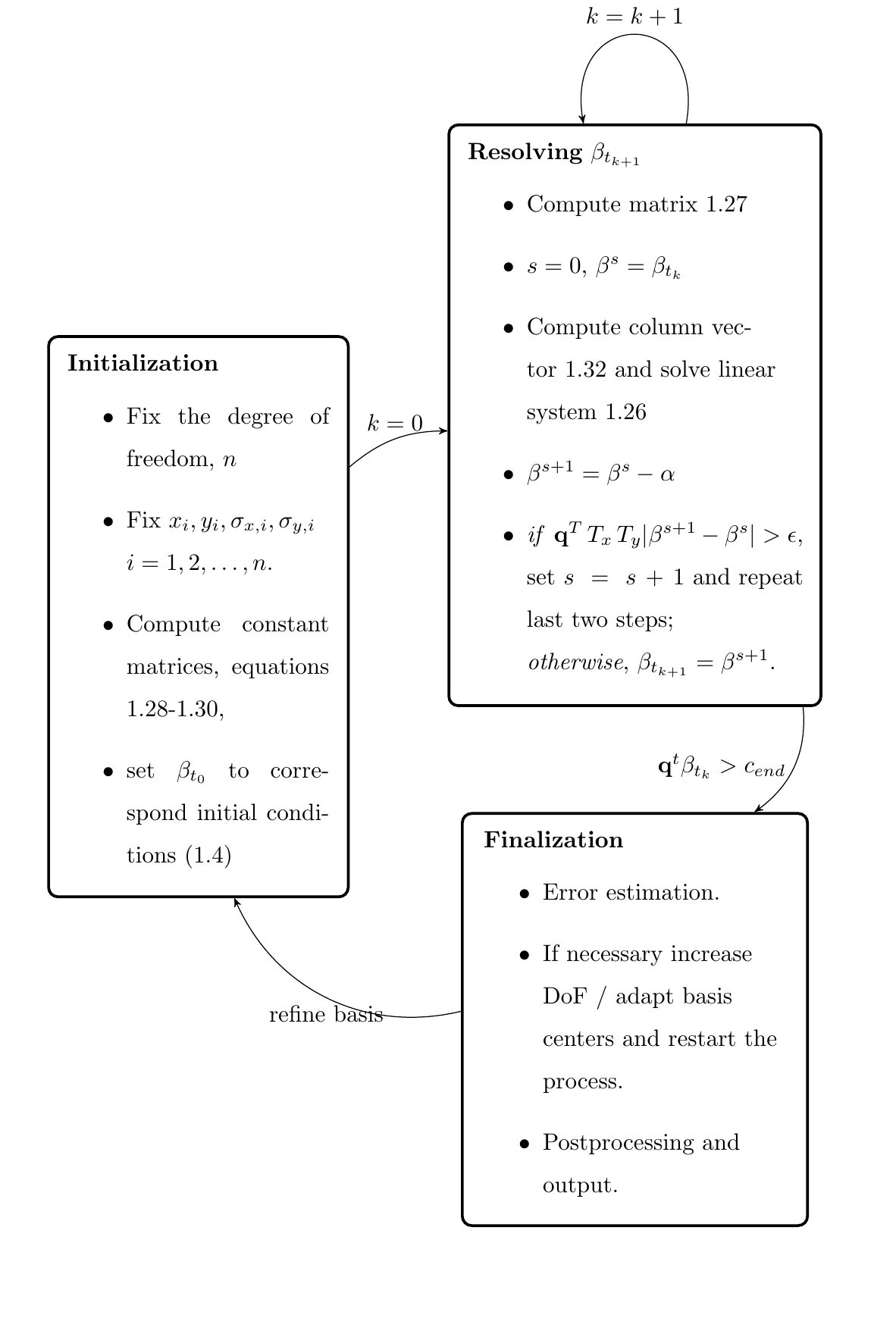}
\end{document}